\documentclass[a4paper]{article}
\usepackage{longtable}

\def\cV{{\mathcal V}}
\def\cA{{\mathcal A}}
\def\cB{{\mathcal B}}
\def\cC{{\mathcal C}}

\def\cL{{\mathcal L}}
\def\cM{{\mathcal M}}
\def\cN{{\mathcal N}}
\def\cP{{\mathcal P}}

\def\cR{{\mathcal R}}

\def\dA{{\dot A}}
\def\dB{{\dot B}}

\def\beq{\begin{equation}}
\def\eeq{\end{equation}}
\def\bea{\begin{eqnarray}}
\def\eea{\end{eqnarray}}
\def\Eee{E_{8(+8)}}

\def\IJ{{\underline{[IJ]}}}
\def\KL{{\underline{[KL]}}}
\def\MN{{\underline{[MN]}}}

\def\asinh{{\rm Arsinh}\,}

\usepackage{amssymb}
\begin{document}

\begin{center}
{\bf\Large The many vacua of gauged extended supergravities}

\bigbreak

{\bf Thomas Fischbacher\\}
\smallbreak
{\em University of Southampton
  School of Engineering Sciences\\
  Highfield Campus\\
  University Road, SO17 1BJ Southampton, United Kingdom\\}
{\small {\tt t.fischbacher@soton.ac.uk}}

\end{center}

\begin{abstract}
\noindent
A novel method is presented which employs advanced numerical
techniques used in the engineering sciences to find and study the
properties of nontrivial vacua of gauged extended supergravity
models. While this method only produces approximate numerical data
rather than analytic results, it overcomes the previous limitation of
only being able to find vacua with large residual unbroken gauge
symmetry groups.  The effectiveness of this method is demonstrated by
applying it to the technically most challenging $D\ge 3$ scalar
potential -- that of ${\rm SO(8)\times SO(8)}$ gauged ${\mathcal
N}=16$ Chern-Simons Supergravity in~$D=3$. Extensive data on the
properties of 99~different vacua (92~of them new) of this model are
given. Furthermore, techniques are briefly discussed which should
allow using this numerical information as an input to the construction
of semi-automatic stringent analytic proofs on the locations and
properties of vacua. It hence is argued that these combined techniques
presumably are powerful enough to systematically map all the
nontrivial vacua of every supergravity model.
\end{abstract}

\section{Introduction}

\noindent This work utilizes techniques that are widely used in 
the Engineering disciplines to solve an old open problem in
supergravity/superstring theory. While these approaches manage to
produce results that have a number of applications considering the
present focus of activity in string theory, the study of supergravity
models itself did not receive nearly as much attention recently as it
did twenty years ago. As the relevant aspects of the problem can be
presented in a self-contained way that requires little more than basic
knowledge in (multi)linear algebra, it is expected that parts of the
program of mapping all supergravity vacua may be carried out by
non-field-theorists. Thus, both a brief gentle introduction to the
physical context (which may be skipped by experts) as well as a
complete specification of the mathematical construction relevant for
the most challenging example -- $\cN=16, D=3$ Chern-Simons
Supergravity with gauge group $SO(8)\times SO(8)$ -- are
given.\footnote{Note that we are not accurate in properly discerning
between~${\rm Spin(8)}$ and $SO(8)$ here. While this model came to be
known as the~$SO(8)\times SO(8)$ gauged model, it is clear that the
double cover must be meant, for the spacetime scalars transform in
spinorial representations.}

\subsection{Motivation and Context}

\noindent Both quantum mechanics and the special theory of relativity are
believed to be based on deep insights into the structure of the
physical world. Yet, under very general assumptions, any attempt to
reconcile them in an unified framework necessarily leads to
relativistic quantum field theory (cf. e.g. \cite{Berestetsky:1982aq,Weinberg:1995mt}).
The perhaps most striking features of quantum field theory are (i)
that it was spectacularly successful in producing quantitative
agreement between theory and experiment hitherto unknown in the whole
history of physics, (ii) that it is a more rigid construction than one
would naively guess and involves a number of `hidden catches' (such as
in the form of anomalies that must cancel) which impose highly
nontrivial restrictions which invalidate many constructions which may
na\"{\i}vely seem to make sense, and finally, (iii) that putting it on
a sound mathematical foundation still remains a highly elusive
problem. Yet, quantum field theory has had some striking successes
both in mathematics and in physics. Apart from e.g. providing a
framework to both explain the origin of and generalize knot theoretic
invariants~\cite{Witten:1988hf}, to effectively compute invariants of
four-manifolds~\cite{Seiberg:1994rs}, and much more, one of the most
spectacular breakthroughs was the (experimentally verified)
unification of the electromagnetic and weak fundamental
forces~\cite{Weinberg:1967tq}. While the framework in which this is
achieved -- nonabelian gauge theory -- is powerful enough to also
allow further unification of the electroweak with the strong force (in
various ways), fundamental obstacles prevent the construction of a
fully unified quantum theory of all fundamental forces, including
gravity. At the heart of this problem lies the observation that the
quanta of gravitational attraction -- gravitons -- differ in important
characteristics from the quanta of the other fundamental forces: while
the source currents of the electromagnetic, weak, and strong
interaction behave in an analogous way under rotations of space, those
of gravity do not. Consequently, a fully unified theory of all forces
would have to involve extensions of spacetime symmetry (i.e. the
Lorentz group) that allow one to overcome this obstacle. Such
extensions do exist, but (due to the strong restrictions imposed by
the framework of quantum field theory) involve a generalization of the
notion of a symmetry group, which, despite being highly un-intuitive,
remarkably seems to be perfectly compatible with the requirements of
quantum field theory.  These `supersymmetry' transformations unify
particles of different spin, and, as such, ultimately even allow
unification of force quanta (bosons of integer spin) with matter
quanta (fermions of half-integer spin).

Even temporarily ignoring quantum mechanics, finding supersymmetric
extensions of general relativity is an interesting problem in
itself. There is a limit on the number of supersymmetry generators
possible due to the restriction that increasing their number unifies
more and more fundamental fields. This makes the construction ever
more rigid until complete unification -- in which all fields are
related to one another through supersymmetry transformations -- is
achieved. Then, both the particle content of the model as well as all
interactions are completely determined by symmetry requirements,
usually up to some discrete choices between alternative possible
constructions, and very few (e.g. one) real parameters. One finds
that, surprisingly, different constructions of supergravity usually
are related to one another by going up and down in the number of
spacetime dimensions. In particular, most supergravity models (though
not the three-dimensional gauged Chern-Simons models we are mainly
concerned with in this work) can be obtained through suitable
compactification of a higher-dimensional ancestor model,
i.e. splitting $D+1$-dimensional spacetime into $3+1$-dimensional
Minkowski space and some $D-3$-dimensional compact space~$\cM$ and
then considering supergravity on such spacetime backgrounds with the
`radius' of~$\cM$ going to zero. By studying supergravity in various
spacetime dimensions, one indeed finds that virtually all models have
a single common ancestor, the (classically) unique 10+1-dimensional
supergravity discovered in~1978 by Cremmer, Julia, and
Scherk~\cite{Cremmer:1978km}.

Compactification of eleven-dimensional supergravity to ten spacetime
dimensions gives supergravity models which are known to be the low
energy limit of ten-dimensional superstring theory. Therefore,
supergravity is intimately linked to superstring theory, and indeed,
it is now widely believed that eleven-dimensional supergravity is the
low energy limit of an elusive eleven-dimensional theory preliminarily
called $M$-Theory\cite{Witten:1995ex}, which can be reduced to each of
the five possible constructions of ten-dimensional superstring theory
and in which two-dimensional membranes rather than strings seem to
play a fundamental role.

Progress in theoretical physics often comes from advances in our
understanding of the role of symmetry principles. It is hence
interesting to observe that -- rather unexpectedly -- new global (and
rather special) symmetries (and hence, due the Noether theorem,
conserved currents) emerge in compactifications of eleven-dimensional
supergravity to lower dimensions. One may conjecture that these can be
traced back to hidden symmetry properties of eleven-dimensional
$M$-Theory that are not well understood yet, but may eventually become
important in the construction of the full
theory~\cite{Cremmer:1997ct,Englert:2003zs,Damour:2002cu,de Buyl:2005mt,
Bergshoeff:2008qd,Bergshoeff:2008xv}.

An important motivation of practical relevance to understand the
detailed structure of supergravity models comes from the `AdS/CFT
conjecture' that some models of supergravity have an equivalent dual
(in the sense of a one-to-one dictionary of notions and statements)
description in terms of conformal field theory in lower-dimensional
space~\cite{Maldacena:1997re}. This duality is exciting as it e.g. may
allow the computation of quantities which are technically difficult
(which may mean: at present impossible) to obtain in a direct approach
by taking a detour through the dual theory.

While a large number of quantum field theories can be constructed that
seem to be free of internal inconsistencies, the rules of the game are
rather limited due to such constraints as anomaly cancellation,
renormalizability, and ghost-freeness. It is therefore not surprising
to see that effects well known from the standard model of particle
physics also play a role in much more general models whose purpose is
more about understanding how quantum field theory works than about
explaining the observed world. In particular, the Higgs
effect~\cite{Higgs} -- spontaneous breaking of a local symmetry due to
a nonzero background of a charged scalar field -- also plays a role in
supergravity models with scalars. What one finds here is that a model
can have different (perturbatively) stable vacuum solutions that
differ by the background values of these scalar fields, and have
markedly different properties of elementary particles. One particular
vacuum solution that recently attracted considerable interest after
25~years of dormancy is the stationary point in the Higgs potential of
maximally supersymmetric four-dimensional supergravity which breaks
the gauged R-symmetry group $SO(8)$ down to $SU(3)\times U(1)$ (which
-- purely coincidentally -- also is the local symmetry underlying
electromagnetism and the strong force). This solution, originally
found by N. Warner~\cite{Warner:1983vz}, seems to be related via
AdS/CFT duality to a 2+1-dimensional field theory model describing the
physics of coincident membranes of 10+1-dimensional
$M$-Theory~\cite{Benna:2008zy,Klebanov:2008vq}, providing another
link\cite{Bagger:2007jr,Gustavsson:2007vu} by which detailed
information about the properties of supergravity vacua may help to
ultimately solve the puzzle of the structure of $M$-Theory.

\subsection{The vacua of gauged extended supergravity}

The existence of scalar field configurations that give stable
alternative vacua of models with extended supergravity was first
demonstrated in~\cite{Breitenlohner:1982jf} for $SO(5)$-gauged
four-dimensional extended supergravity. From the perspective of
nontrivial vacuum solutions, the most challenging and most interesting
supergravity models are those which -- in $D$~dimensions -- feature a
hidden global $E_{11-D(11-D)}$ symmetry. In~$D=3$, this global
symmetry group is the (maximal split real form of the) largest
exceptional Lie group $E_8$. These models may be deformed in such a
way that different subgroups of this exceptional global symmetry get
promoted to a gauge group. Retaining supersymmetry then mandates the
introduction of a potential for the scalar fields. Usually (but not
always), setting all scalar fields to zero gives a stable vacuum
solution which then has the largest residual unbroken symmetry. Other
stationary points, which usually exist (sometimes in large number),
are not true extrema but saddle points in the potential, may or may
not correspond to stable vacua, and spontaneously break the gauge
symmetry to smaller gauge groups.

\section{The scalar potential of gauged ${\mathcal N}=16$ $D=3$
         Supergravity}
\label{sec:n16}

\subsection{On the choice of the example}

The ${\rm SO(8)\times SO(8)}$ gauged ${\mathcal N}=16$ $D=3$ model was
chosen as the primary example to explain the method for these reasons:

\begin{itemize}

\item From the purely technical perspective, the scalar potential 
of this model is singled out as the most challenging one in 
$D\ge 3$: it lives on the 128-dimensional octooctonionic 
plane ${\rm E_{8(8)}/(Spin(16)/\mathbb{Z}_2)}$ and does not show any 
of the simplifications that occur for other gauge groups not 
contained in ${\rm Spin(16)/\mathbb{Z}_2\subset E_{8(8)}}$.

\item Despite the complexity of this potential, detailed data 
(including mass spectra) are available on a number of nontrivial
vacua: this allows independent validation of results.

\item As the three-dimensional maximal gauged models have been 
discovered more recently than higher-dimensional ancestors, literature
available on them uses more systematized concepts which also
generalize to the higher-dimensional cases.

\item Folk lore says that vacua in higher-dimensional gauged maximal 
supergravity models have counterparts in lower dimensions. We
therefore naturally expect both the vacuum structure in $D=3$ to be
richer than in higher dimensions, as well as to give clues about
possible higher-dimensional ancestors of newly discovered vacua.

\end{itemize}

\subsection{General structure of the problem}

The full Lagrangian of this model is given
in~\cite{Nicolai:2001sv}. It is not repeated here (although this would
be desirable for the sake of completeness) as this would require us to
introduce conventions for $3+1$-dimensional spacetime vector and
spinor indices which would inevitably clash with our extensive
conventions on subgroup index alphabets. In this work, we are (almost)
exclusively concerned with the scalar potential~$\cL_{\rm Vac}$ as
well as a related auxiliary quantity, which (in this context) will be
called the `misalignment' $\cM^{I\dA}$:

\beq
\begin{array}{lcl}
\cL_{\rm Vac}/(g^2e)&=&\frac{1}{8}\left(A_1^{IJ}A_1^{IJ}-\frac{1}{2}A_2^{I\dA}A_2^{I\dA}\right)\\
\cM^{I\dA}&=&3\,A_1^{IM}A_2^{M\dot A}-A_2^{I\dot B}A_3^{\dot A\dot B}
\end{array}
\eeq

These expressions are defined on the 128-dimensional octooctonionic
plane $\cP$ as follows: the matrices $A_1, A_2, A_3$ of respective size
$16\times 16$, $16\times 128$, and $128\times 128$ are specific 
linear combinations (details to be given further down) of entries of 
the so-called T-tensor, which in this case is a $248\times 248$
matrix of the form $T=\cV^T\Theta\cV$, where the (usually sparsely 
occupied) $248\times 248$ matrix $\Theta$ is a constant that describes 
the embedding of the gauge group -- here ${\rm SO(8)\times SO(8)}$ 
-- into $E_{8(8)}$. $\cP$ can be understood as being embedded into
the 248-dimensional manifold $E_{8(8)}$ which itself is a subgroup of 
the double cover of the noncompact real form ${\rm SO(128,120)}$ of 
the group of rotations in 248 dimensions. There is a $1{:}1$ 
correspondence between points on $\cP$ and matrices $\cV$ of the form
$\cV=\exp\left(c^A G^{(128)}_A\right)$ with $c^A$ being an entry of
a 128-dimensional vector of coordinates on $\cP$ and $G^{(128)}_A$ 
the corresponding $248\times 248$ generator matrix of $E_{8(8)}$.
The potential $\cL_{\rm Vac}$ has a local maximum at the origin, 
$c^A=0$, which retains maximal symmetry. While there seem to be no 
further true extrema, it is known that there nevertheless are 
additional stationary (saddle) points which in some cases correspond
to meaningful (perturbatively) stable physical backgrounds (vacua).
The saddle points $d \cL/d c^A = 0$ can be alternatively characterized 
by the vanishing of the $16\times 128$ matrix $\cM^{I\dA}$.

\subsection{Detailed conventions}

For the sake of easy reproducibility of results, conventions are given
in full detail. We generally use the Einstein summation convention:
any index which occurs twice in a product of tensors is implicitly
being summed over. For Spin(N) groups, we do not discern between upper
and lower indices.

Conventions for indices denoting different group representations are:

\begin{tabular}{|l|l|}
\hline
Indices&range (representation)\\
\hline
$I,J,K,\ldots$& $1\ldots16$ (Spin(16) vector)\\
$\IJ,\KL,\ldots$& $1\ldots120$ (Spin(120) adjoint)\\
$A,B,C,\ldots$& $1\ldots128$ (Spin(16) spinor)\\
$\dot A,\dot B,\dot C,\ldots$& $1\ldots128$ (Spin(16) co-spinor)\\
$\cA, \cB, \cC,\ldots$&$1\ldots248$ ($E_{8(8)}$ fundamental/adjoint)\\
$i,j,k,\ldots$&$1\ldots 8$ (Spin(8) vector)\\
$\alpha,\beta,\gamma,\ldots$&$1\ldots 8$ (Spin(8) spinor)\\
$\dot\alpha,\dot\beta,\dot\gamma,\ldots$&$1\ldots 8$ (Spin(8) co-spinor)\\
\hline
\end{tabular}

We generally list only the nonzero components of all tensors. At the
heart of the construction lies the ${\rm Spin(8)}$ Clifford algebra,
given as a set of 8 $8\times 8$ matrices $\gamma^i$ that satisfy

\beq
\begin{array}{lcl}
\gamma^i{}_{\alpha\dot\gamma}\gamma^j{}_{\alpha\dot\delta} +
\gamma^j{}_{\alpha\dot\gamma}\gamma^i{}_{\alpha\dot\delta} &=&
 2\delta^{ij}\delta_{\dot\gamma\dot\delta}\\
\gamma^i{}_{\alpha\dot\gamma}\gamma^j{}_{\beta\dot\gamma} +
\gamma^j{}_{\alpha\dot\gamma}\gamma^i{}_{\beta\dot\gamma} &=&
 2\delta^{ij}\delta_{\alpha\beta}\\
\gamma^i{}_{\alpha\dot\gamma}\gamma^i{}_{\beta\dot\delta} +
\gamma^i{}_{\beta\dot\gamma}\gamma^i{}_{\alpha\dot\delta} 
&=& 2\delta_{\alpha\beta}\delta_{\dot\gamma\dot\delta}\\
\gamma^i{}_{\alpha\dot\gamma}\gamma^i{}_{\beta\dot\delta} +
\gamma^i{}_{\alpha\dot\delta}\gamma^i{}_{\beta\dot\gamma} 
&=& 2\delta_{\alpha\beta}\delta_{\dot\gamma\dot\delta}\\
\gamma^i{}_{\alpha\dot\gamma}\gamma^j{}_{\alpha\dot\delta} +
\gamma^i{}_{\alpha\dot\delta}\gamma^j{}_{\alpha\dot\gamma} 
&=& 2\delta^{ij}\delta_{\dot\gamma\dot\delta}\\
\gamma^i{}_{\alpha\dot\gamma}\gamma^j{}_{\beta\dot\gamma} +
\gamma^i{}_{\beta\dot\gamma}\gamma^j{}_{\alpha\dot\gamma} 
&=& 2\delta^{ij}\delta_{\alpha\beta}
\end{array}
\eeq

Our detailed conventions on $\gamma^i{}_{\alpha\dot\gamma}$ will
become important later on, for different reasons. It is possible to
satisfy these conditions by $8\times 8$ matrices which each have one
single entry $\pm 1$ per row and column. This following choice
reproduces the conventions given in \cite{Green:1987sp}: we concisely
list all 64~nonzero entries $\gamma^i{}_{\alpha\dot\gamma}$ in the 
form $PQR_\pm$, meaning $\gamma^{i=P}{}_{\alpha=Q\,\dot\gamma=R}=\pm1$:

\beq
\begin{array}{llllllll}
118_{+}&127_{-}&136_{-}&145_{+}&154_{-}&163_{+}&172_{+}&181_{-}\\
212_{+}&221_{-}&234_{-}&243_{+}&256_{+}&265_{-}&278_{-}&287_{+}\\
315_{+}&326_{-}&337_{+}&348_{-}&351_{-}&362_{+}&373_{-}&384_{+}\\
413_{+}&424_{+}&431_{-}&442_{-}&457_{-}&468_{-}&475_{+}&486_{+}\\
514_{+}&523_{-}&532_{+}&541_{-}&558_{+}&567_{-}&576_{+}&585_{-}\\
616_{+}&625_{+}&638_{+}&647_{+}&652_{-}&661_{-}&674_{-}&683_{-}\\
717_{+}&728_{+}&735_{-}&746_{-}&753_{+}&764_{+}&771_{-}&782_{-}\\
811_{+}&822_{+}&833_{+}&844_{+}&855_{+}&866_{+}&877_{+}&888_{+}
\end{array}
\eeq

From these eight matrices, we form the real $128\times 128$ matrices
$\Gamma^I_{A\dB}$ of the Clifford algebra of 16-dimensional Euclidean
space via the embedding\\
${\rm Spin(8)\times Spin(8)\subset
Spin(16)}$. It generally makes sense to define a set of index
splitting/embedding matrices for all of which we will use the name $H$
and which are discerned by the types of indices they carry. For these,
we use the conventions given below. Here, we encounter an annoying
issue that can be traced back to the widespread yet conceptually
awkward convention to use indices for a $D$-dimensional representation
in the range $1\ldots D$ rather than $0\ldots(D-1)$. In order to deal
with this, we define $\cR(p,q):=(p-1)\cdot 8+(q-1)+1$. Thus, we have:
\beq
\begin{array}{lclclcl}
H^I_i&=&\delta^{I,i}&\qquad&H^I_{\bar i}&=&\delta^{I,\bar i+8}\\
H^A_{\alpha\beta}&=&\delta_{A,\cR(\alpha,\beta)}&&
H^A_{\dot\alpha\dot\beta}&=&\delta_{A,64+\cR(\dot\alpha,\dot\beta)}\\
H^{\dot A}_{\alpha\dot\beta}&=&\delta_{\dot A,\cR(\alpha,\dot\beta)}&&
H^{\dot A}_{\dot\gamma\delta}&=&\delta_{\dot A,64+\cR(\dot\gamma,\delta)}\\
H^\cA_B&=&\delta^{\cA,B}&\qquad&H^\cA_{\IJ}&=&\delta^{\cA,128+\IJ}
\end{array}
\eeq

Additionally, we use the convention
\beq
\delta^{i_1\ldots i_n}_{k_1\ldots k_n} = (1/n!)\cdot\sum_{p\in\pi_n} (-1)^{{\rm sign}(p)} \delta^{i_1}_{k_{p_1}}\cdots  \delta^{i_n}_{k_{p_n}}
\eeq
so that this `generalized Kronecker delta' can be used as an anti-symmetrizing projector.

Then,
\begin{equation}
\begin{array}{lcl}
\Gamma^{I}_{A\dot A}&=&\phantom+H^I_i H^A_{\alpha\beta} H^{\dot A}_{\gamma\dot\delta}\delta_{\alpha\gamma}\gamma^i_{\beta\dot\delta}
+H^I_i H^A_{\dot\alpha\dot\beta} H^{\dot A}_{\dot\gamma\delta}\delta_{\dot\alpha\dot\gamma} \gamma^i_{\delta\dot\beta}\\
&&+H^I_{\bar i} H^A_{\alpha\beta} H^{\dot A}_{\dot\gamma\delta}\delta_{\beta\delta}\gamma^{\bar i}_{\alpha\dot\gamma}
-H^I_{\bar i} H^A_{\dot\alpha\dot\beta} H^{\dot A}_{\gamma\dot\delta}\delta_{\dot\beta\dot\delta}\gamma^{\bar i}_{\gamma\dot\alpha},\\
\Gamma^{IJ}_{AB}&=&\delta^{IJ}_{KL}\Gamma^{K}_{A\dot C}\Gamma^{L}_{B\dot C}.
\end{array}
\end{equation}

Considering efficient numerical computations involving explicit sparse
matrix representations, these particular conventions have a number of
remarkable highly desirable properties which do not hold in general
but will come to play an important role to make numerical computations
fast (this is our first important trick!), such as in particular:
\beq\label{fastgamma}
\Gamma^I{}_{A\dot C}\Gamma^J{}_{D\dot C}\Gamma^K{}_{D\dot E}\Gamma^L{}_{B\dot E}\delta^{IJKL}_{MNPQ} = \Gamma^M{}_{A\dot C}\Gamma^N{}_{D\dot C}\Gamma^P{}_{D\dot E}\Gamma^Q{}_{B\dot E}
\eeq
\beq
\Gamma^I{}_{A\dot C}\Gamma^J{}_{D\dot C}\Gamma^K{}_{D\dot E}\delta^{IJK}_{MNP} = \Gamma^M{}_{A\dot C}\Gamma^N{}_{D\dot C}\Gamma^P{}_{D\dot E}.
\eeq

As usual, we further define:
\beq
\Gamma^{KL}_{AB}:=\delta_{IJ}^{KL}\Gamma^I{}_{A\dot C}\Gamma^J{}_{B\dot C}
\eeq

Before we can proceed to define the 248 sparse $248\times 248$
matrices $(G^{(E8)}_\cA){}^\cC{}_\cB$ (the generators of the group $\Eee$),
we need further conventions that extract the adjoint index $\IJ$ from 
a pair of vector indices $I,J$. We also use the name $H$ for this form 
of index projection, in detail:
\beq
\begin{array}{lcl}
H^{\IJ}_{KL}&=&\phantom+\delta_{M=\IJ,M=(K-1)\cdot 16+L-K(K+1)/2}\\
&&-\delta_{M=\IJ,M=(L-1)\cdot 16+K-L(L+1)/2}
\end{array}
\eeq
and likewise for $H_{\IJ}^{KL}$. Also, we define
$H^\cA_{IJ}:=H^{\cA}_{\IJ}H^{\IJ}_{IJ}$.

With these, we first introduce generators for the group 
${\rm Spin(16)}$ of rotations in 16 dimensions, in the 
adjoint representation:
\beq
\left(G^{(\rm Spin(16))}_\IJ\right)^\KL{}_\MN=
-H_\IJ^{IJ}H^{\KL}_{KL}H_{\MN}^{MN}\delta^{IJ}_{I'J'}\delta^{KL}_{K'L'}\delta^{I'K'}\delta_{L'J'}^{MN}
\eeq

Then, the $(G^{(E8)}_\cA){}^\cC{}_\cB$ are given by:
\beq
\begin{array}{lcl}
(G^{(E8)}_\cA){}^\cC{}_\cB&=&H^\cA_{\IJ}H^\cB_{\KL}H_\cC^{\MN}\left(G^{(\rm Spin(16))}_\IJ\right)^\KL{}_\MN\\
&&+\frac{1}{4}\Gamma^{IJ}_{AB}\left(H^\cA_{IJ}H^\cB_AH_\cC^B+H^\cA_BH^\cB_{IJ}H_\cC^A-H^\cA_A H^\cB_B H_\cC^{IJ}\right)
\end{array}
\eeq

Finally, we need the gauge group embedding tensor $\Theta_{\cA\cB}$,
which for the particular case ${\rm SO(8)\times SO(8)}$ has the form:
\beq
\Theta_{\cM\cN}=H^{\IJ}_{\cM}H^{\KL}_{\cN}H_{\IJ}^{IJ}H_{\KL}^{KL}\left(H_I^{\bar i}H_J^{\bar j}H_K^{\bar k}H_L^{\bar l}\delta_{\bar i\bar j}^{\bar k\bar l}-H_I^iH_J^jH_K^kH_L^l\delta_{ij}^{kl}\right)
\eeq

Due to the special role of vector/scalar duality in three dimensions,
the constraints on possible gauge groups are less stringent than in
higher-dimensional maximal supergravity: Any subgroup $H\subset
E_{8(+8)}$ for which the corresponding embedding tensor does not carry
a contribution in the 27000-dimensional irreducible representation of
$E_8$ can be chosen as gauge group. Apart from the non-compact forms
(${\rm SO(p,8-p)\times SO(p,8-p)}$ for $1\le p\le 7$), which have
analoga in higher dimensions such as ${\rm SO(p,8-p)}$ gauged
${\cN=8}$ supergravity in $D=4$, there are a number of additional
possibilities here that include exceptional gauge
groups~\cite{Nicolai:2001sv}, such as e.g. $G_{2(2)}\times F_{4(4)}$
or $E_{7(-5)}\times SU(2)$.

Given the $(G^{(E8)}_\cA){}^\cC{}_\cB$, the explicit expressions 
for the matrices $A_1, A_2, A_3$ as functions of 128 coordinates 
on the octooctonionic plane $c^A$ are:
\beq
\begin{array}{lcl}
\cV^{\cM}{}_\cA &=& \exp\left(c^A H_A^\cA G^{(E8)}_\cA\right)^\cM{}_\cA\\
T_{\cA\cB}&=&\cV^\cM{}_\cA \Theta_{\cM\cN} \cV^\cN{}_\cB\\
\eta^{\cA\cB}&=& H^\cA_A H^\cB_B\delta{AB} - H^\cA_\IJ H^\cB_\KL \delta^\IJ_\KL\\
\theta&=&\frac{1}{248}\eta^{\cM\cN}\Theta_{\cM\cN}\\
A_1^{IJ}&=&\frac{8}{7}\theta\delta_{IJ}+\frac{1}{7}T_{\cA\cB} H^\cA_{\underline{[IK]}}H^\cB_{\underline{[JK]}} H^{\underline{[IK]}}_{IK} H^{\underline{[JK]}}_{JK}\\
A_2^{I\dot A}&=&-\frac{1}{7} T_{\cA\cB} H^\cA_{\underline{[IJ]}} H^{\underline{[IJ]}}_{IJ} H^\cB_{A}\Gamma^J_{A\dot A}\\
A_3^{\dot A\dot B}&=&2\theta\delta_{\dot A\dot B}+\frac{1}{48}\Gamma^{IJKL}_{\dot A\dot B}T_{\cA\cB} H^\cA_{\underline{[IJ]}}H^\cB_{\underline{[KL]}} H^{\underline{[IJ]}}_{IJ} H^{\underline{[KL]}}_{KL}
\end{array}
\eeq

Finally the scalar potential is then given as:
\beq
\cL_{\rm Vac}/(g^2e)=\frac{1}{8}\left(A_1^{IJ}A_1^{IJ}-\frac{1}{2}A_2^{I\dA}A_2^{I\dA}\right)
\eeq

\section{The Method}

Previous approaches towards finding stationary points of supergravity
potentials on high-dimensional scalar manifolds involved analytic
treatments based on a restriction of the full problem to
low-dimensional submanifolds invariant under some (usually large)
chosen subgroup of the gauge
group~(e.g. \cite{Warner:1983du,Warner:1983vz,Hull:1984ea}). While
this is a powerful approach, especially in conjunction with symbolic
algebra tailored to the task~\cite{Fischbacher:2002fr}, a fundamental
problem is the inability to find stationary points which do not lie on
any low-dimensional submanifold accessible in such a way. Hence, a
complementary numerical approach that is able to yield information
about otherwise inaccessible solutions is highly desirable.

One fundamental objection against the idea of tackling the problem
numerically is that stationary points are not isolated but lie on a
manifold that is invariant under rotations in the gauge group, hence
we should expect to obtain their position in general form, which is a
mostly meaningless collection of coordinates. There are, however,
simple numerical techniques to address this problem and ensure that
stationary points are automatically rotated into a useful form. This
will be explained at the end of this section. 

The basic ingredients of the numerical approach are:

\begin{enumerate}

\item A real-valued function $P$ that can be evaluated fast to high accuracy 
      which provides information about how much a given point violates the 
      stationarity condition. (Using the misalignment tensor $\cM$, 
      this requires fast computation of the tensor $A_3$, made possible
      by equation~(\ref{fastgamma}).)

\item A fast method to evaluate the gradient of $P$.

\item An efficient numerical algorithm to find minima of $P$
      in constrained space. (Practically all such algorithms
      utilize the gradient of $P$.)

\item Sufficient numerical precision to allow the extraction of useful data.

\item Post-processing of raw data to obtain information about particle
       mass spectra.

\item A reduction step that brings a generic point on the submanifold
      of equivalent vacua to a simple coordinate representation by
      applying a suitable gauge group rotation.

\end{enumerate}

\subsection{The quasipotential $P$}

As it stands, the potential $\cL_{\rm vac}$ is not well suited for the
purpose of performing a numerical search for stationary points, all of
which (except for the trivial one at the origin) are saddle
points. Therefore, we instead base our search on the function:
\beq
\tilde P=\cM_{I\dA} \cM_{I\dA}
\eeq

Here, we have $\tilde P\ge 0$ and $\tilde P=0$ precisely for the
stationary points of $\cL_{\rm vac}$. While it often is not a good
idea to recast the problem of numerically solving an equation as an
optimization problem of the form $([{\rm LHS}]-[{\rm RHS}])^2=\rm
min!$, this works sufficiently well for our purposes here. However, as
it stands the function $\tilde P$ also is not very well suited for
numerical optimization, as the degree of violation of the stationarity
condition measured by $\tilde P$ depends roughly exponentially on the
coordinate distance of the point in the octooctonionic plane from the
origin. Consequently, optimization algorithms that iteratively try to
reduce the objective function will encounter extremely steep gradients
close to extremely narrow `golf hole style' solutions, which causes the
optimizer to `wildly bounce around' in the allowed parameter
regime. While it is precisely this behaviour that may occasionally
succeed in producing solutions that are otherwise unlikely to arise,
one would nevertheless generally prefer a somewhat more well-behaved
function. For the analysis presented here, the choice was
\beq
P = \left(\asinh \tilde P\right)\left(c^A c^A\right)^\rho
\eeq

Without the extra factor $\{{\rm radius}\}^{2\rho}$, the inverse
hyperbolic sine of the exponential would give a potential which for
generic configurations is roughly linear in the distance from the
origin -- up to extra dips corresponding to nontrivial
solutions. However, the `funnel-like' shape would make the region
around the origin so attractive that it becomes rather difficult to
find solutions lying far out. Setting the constant $\rho$ to about
$-0.5$ turns the funnel into a (more or less) smooth plateau with a
very broad minimum. By changing this parameter, one can tune the
optimization algorithm to spend more time in regions of varying
distance from the origin, and hence systematically search for
solutions lying in a certain distance region.

\subsection{Fast gradients through sensitivity analysis}

At about the same time when N.~Warner worked on the structure of the
four-dimensional cousin of the model studied
here~\cite{Warner:1983du,Warner:1983vz} in Supergravity, Speelpenning
-- working in computer science -- presented an algorithmic approach
that allows the fast evaluation of gradients for scalar functions of
many parameters~\cite{Speelpenning1980} that also can be used to great
effect for the analysis of supergravity potentials. Essentially, he
showed that, given a differentiable real-valued function $f:
\mathbb{R}^n\rightarrow \mathbb{R}$ of~$n$ parameters, then a computer
program that computes the value of~$f$ at a specific point $P$ within
time~$T$, can be mechanically translated to a computer program that
computes \emph{to high accuracy} the $n$ entries of $\nabla f$ at $P$
within a time bounded by~$5T$, \emph{regardless of the number 
of parameters~$n$}!

This evidently is far more powerful than the naive approach to compute
the gradient from finite differences, which would require at least a
computation time of $(n+1)\cdot T$ and not produce highly accurate
results.

The price that has to be paid when using Speelpenning's method is that
(i) all intermediate results in the computation have to be retained
(rather than overwritten) as~$f$ is evaluated, (ii) for each
intermediate numerical quantity, memory space has to be reserved to
hold another number, and (iii) at the end of the first stage of the
`fast gradient' algorithm, which is the computation of~$f$ at the
given point, the sequence of operations that were executed to produce
this number must be known.

The third requirement means that, for computations that involve
decisions on the magnitude of intermediate quantities (e.g. a number
of iterations depending on when a certain tolerance level is reached),
information on how these decisions were made during the computation
has to be retained. Therefore, automated program transformers that map
computer code for~$f$ to fast computer code for~$\nabla f$ produce
code which records an execution `tape' data structure that contains
information on all these decisions in the computation of~$f$. If the
computation of~$f$ is structurally simple enough -- as is fortunately
the case with supergravity potentials and the quasipotentials
introduced to measure deviation from the stationarity condition --
then it is often advantageous not to use an automated program
transformer (such as e.g. ADOLC~\cite{Griewank1996}), but to instead
manually re-write the code in such a way that it also offers fast
gradients, as this allows tighter code and easily avoids the (slightly
awkward) `tape' data structure.

Briefly stated, the idea underlying the fast computation of the
gradient of~$f$ at the point~$X=(x_1,\ldots,x_n)$ is to first compute
$z=f(X)$ and associate with every intermediate quantity~$y$ an extra
quantity~ $\bar y$, which gives $\delta z/\delta y$, i.e. the change
of the final result relative to an infinitesimal change of this
intermediate value~$y$: \emph{if, after the determination of~$y$, the
computation continued not with this value~$y$, but instead
with~$y+\delta y$, how much would this change the result~$z$}? 
Starting with the final computation that produced~$z$, we can then use
the chain rule to backpropagate sensitivities to earlier and earlier
intermediate results. Treating input parameters~$x_k$ in the same way
as the intermediate quantities by also associating output
sensitivities~$\bar x_k=\delta z/\delta x_k$ to these, we eventually
obtain the rates of change of the output with each of the inputs,
\emph{which is just the gradient}. The backpropagation algorithm
hence works as follows:

\begin{enumerate}

\item For every input and intermediate quantity, and also for the result,
      allocate enough memory to hold two floatingpoint numbers.
      Initialize all of them to~\texttt{0.0}. 

\item Load the input quantity value cells $x_n$ with the input parameters.

\item Perform the computation of~$f$, remembering sufficient information 
      to later retrace in reverse all numerical operations
      performed. This gives the result~$z$. None of the cells holding
      associated sensitivities are touched.

\item Set the content of the memory cell holding the associated 
      sensitivity~$\bar z$ of the result~$z$ to~\texttt{1.0}.
      ($\delta z/\delta z=1$).

\item Retrace all numerical computation steps in reverse.
      At each step, the output sensitivity of its computational
      result~$y_m$ is known: $\bar y_m=\delta z/\delta y_m$. Update the
      sensitivities of the arguments of the present step that produced~$y_m$
      according to these rules:

  \begin{itemize}
   \item \emph{if it is an addition} $y_m=y_a+a_b$, then increase the number
         stored in the cell holding~$\bar y_a$ by~$\bar y_m$, and also increase
         the number stored in the cell holding~$\bar y_b$ by~$\bar y_m$.

   \item \emph{if it is a multiplication} $y_m=y_a\cdot y_b$, then increase the number
         stored in the cell holding~$\bar y_a$ by~$\bar y_m\cdot y_b$, and also
         increase the number stored in the cell holding~$\bar y_b$ by~$\bar y_m\cdot y_a$.

   \item \emph{if it is a function application} $y_m=g(y_k)$, then increase the number
         stored in the cell holding~$\bar y_k$ by~$\bar y_m\cdot (\partial g/\partial y_k)(y_k)$.

  \end{itemize}

  If one of the summands(/factors) in an addition(/multiplication)
  operation is a constant, then this does not come with an associated
  sensitivity, hence we omit the corresponding update step: we are not
  interested in how the result would change if we changed the fixed
  constants in the computational algorithm. Note that this situation
  is covered by the `function application' rule: if e.g. 
  $y_{20}=5\cdot y_{17}$, then we can take $g=(x\mapsto 5\cdot x)$,
  and hence have to update $\bar y_{17}$ by increasing it by~$5\cdot y_{20}$.
  The `function application' rule also handles divisions, as we can regard 
  the fraction $p/q$ as the product $p\cdot h(q)$ with~$h=(x\mapsto 1/x)$.
  When manually re-writing existing code for backpropagation,
  it is occasionally useful to put in modifications that allow a comparison
  between an individual intermediate backpropagated sensitivity and the change
  of the result produced by redoing the computation with an artificially
  introduced small change to the corresponding intermediate quantity.
  This allows easy validation of the new code and simplifies locating bugs.

\end{enumerate}

It may be useful to convince oneself how backpropagation works by
studying a simple example that is structurally similar to the
computation of a supergravity potential, such as
\beq
v=\exp x,\quad t=v^2,\quad a=1/7\cdot t,\quad z=(\asinh a^2)\cdot\left(x^2\right)^{0.4}
\eeq

A more detailed description of backpropagation can be found
e.g. in~\cite{Griewank1989}. Considering the associated semantic aspects,
an interesting question is what would have to be done in order to
extend a given programming language with a `backpropagation operator'
that maps a function to a fast gradient function and allows
nesting. This has been studied in~\cite{Pearlmutter2008}.

It makes sense to implement the functions to compute the supergravity
potential, as well as the quasipotential and its gradient, in the C
programming language, and then interface this to the high-level
scripting language Python in order to do the optimizing part of the
problem with a readily existing efficient numerical optimizing library
routine. The Python programming language is an attractive choice here,
as a number of useful libraries for tasks such as numerical
optimization are readily available due to its high popularity in the
Engineering Sciences. In this work, we use the \texttt{scipy.optimize.fmin\_l\_bfgs\_b} 
function~\cite{Byrd1995} from the `Scientific Python' package~\cite{scipy}.

\subsection{Other ingredients}

A number of minor additional technicalities are used in the fast
numerical determination of supergravity vacua:

\subsubsection{Gamma matrices}

The computationally relevant Spin(16) Gamma
matrices~$\Gamma^I,\,\Gamma^{IJ},\,\Gamma^{IJK},\,\Gamma^{IJKL}$ all
share the property that they are real $128\times128$ matrices that
have a single nonzero entry per row and per column which is
$\pm1$. Due to~(\ref{fastgamma}), they allow a memory-efficient compact
code representation that stores each of the $\Gamma^I$ as a string of
128 signed 8-bit characters where the $k$-th entry gives both the
sign (1 bit) as well as the index (7 bit) of the single nonzero entry
in the $k$-th column of $\Gamma^I$. As we also need the transpose of
$\Gamma^I$, this amounts to a total memory requirement of
$16\cdot128\cdot 2 = 4096$ byte, as well as some simple command logic
to string subsequent lookups and keep track of the sign when dealing
with antisymmetrized higher powers of $\Gamma$ matrices. These
operations are fast, as they easily fit into the CPU's L1 caches.

\subsubsection{Fast matrix exponentiation}

The sensitivity backpropagation techniques strongly favor a simple
algorithm that is reasonably fast to numerically compute $E_8$~group
elements by exponentiating $E_8$~generators. According to the overview
paper~\cite{Moler2003}, the method which seemingly also was used in
early versions of MATLAB~\cite{matlab} suggests itself here: in order
to compute~$\exp M$, we first compute $Q=2^{-n}\cdot M$ for $n$
sufficiently large to make the numerical computation of~$\exp Q$ via
the Taylor series expansion reasonably fast (i.e. $\|Q\|_2< 1$), and
then square the result $n$~times, using~$\exp(A)^2=\exp(2A)$.

\subsubsection{Search parameters: Initial point and the parameter $\rho$}

As the numerical approach presented here only allows `fishing' for
vacua, the question arises how to effectively choose parameters to
obtain a large number of different vacua. Considering the structure of
the potential, one could start by choosing all 128 spinor components
as equidistributed points within a~$[-A;A]^{128}$ cube. This, however,
enormously disfavors points close to hyperface midpoints relative to
points close to hyperdiagonals. A `fairer' strategy to choose an
initial point seems to be to first choose a random point on the
circle~$S^2$ and then use its coordinates as relative weights for two
random points on~$S^{64}$. While this approach is indeed viable and
produces a number of nontrivial solutions, it must be pointed out that
one should not resort to a single strategy here. In particular, most
of the stationary points at a large distance from the origin which are
presented here were found by (i) using equidistribution on an
128-dimensional cube for the initial point, and (ii) removing both the
exponent~$\rho$ as well as using the quasipotential function~$\tilde P$ 
instead of~$P$.

\subsubsection{Mass spectra}

When numerically determining matrix eigenvalues to obtain particle
mass spectra, having easy access to a powerful eigenvalue function
from a well-maintained software library is very helpful. The
`Scientific Python'~\cite{scipy} package (which for many purposes can
be regarded as a viable alternative to MATLAB) provides
the~\texttt{scipy.linalg.eig} function that can be put to good use
here.

\subsubsection{Beautifying the result}

If the numerical search for a vacuum succeeds, it will return a
stationary point which differs from what one would recognize as a
`nice' presentation of the vacuum by an arbitrary rotation in the
gauge group. So, the question arises how `wild' numbers produced by
numerical optimization that describe a vacuum can be used to arrive at
a `tame' description. One simple strategy is to again use the same
numerical tricks and techniques employed to find the vacuum in the
first place (i.e. sensitivity analysis, fast matrix exponentiation,
efficient numerical optimization, etc.) in order to find a linear
combination of the generators of the gauge group which, after
embedding in~$E_8$ and exponentiation to obtain the corresponding
rotation, manages to minimize the objective function of the
128-dimensional coefficient vector~$c$
\beq
F(c)=\sum_{A=1}^{128} -\left(c^A\right)^4.
\eeq
The (very simple) idea here is that, while $\cos^2 \phi+\sin^2\phi$ is
rotationally invariant, the quantity $\cos^4 \phi+\sin^4\phi$ takes on
a maximum when~$\phi$ is a rotation that aligns the coordinate axes
with themselves. So, this optimization favors setting as many vector
entries to zero as possible by maximizing entries which already were
large in the beginning. This solves the problem sufficiently well for
our purposes, but the strategy may easily be refined to instead
maximize e.g. the coefficients of the $N$-forms contained in the ${\rm
Spinor}\otimes{\rm Spinor}$ representation of the diagonal~$SO(8)$
subgroup of the gauge group, or similar constructions.

\section{New Results}

The code that has been used to identify all the new vacua presented in
this work can be obtained by downloading the source code archive of
this paper from {\tt arXiv.org}\footnote{The corresponding URL is {\tt
http://arxiv.org/e-print/0811.1915}}. The high-level Python script
\texttt{e8\_vacuum} from this source can find, beautify, analyze, and
automatically typeset data describing nontrivial stationary points in
the scalar potential. As the method is not intrinsically exhaustive,
but will only produce a `random' solution (or -- occasionally -- not
converge at all), the following list of program-generated solutions is
expected not to be complete even over the limited range
investigated. Some of the vacua already were known from earlier work
by the author and are listed again for cross-validation purposes of
the numerical approach against symbolic group-theory based methods.

In these tables, the (approximate) `length' of the Spin(16)
spinor~$R=\left(c^Ac^A\right)^{1/2}$ is given together with the
(approximate) value~$\Lambda$ of the potential at the stationary
point, the number of unbroken (left and right) supersymmetry
generators, as well as the dimension of the residual unbroken gauge
group~$H\subset G=SO(8)\times SO(8)$. The locations of these vacua are
given by listing the nonzero coefficients of the 128-dimensional
spinor as a set of
contributions~$\mbox{\{coefficient\}}[\alpha\beta]$,
resp.~$\mbox{\{coefficient\}}[\dot\alpha\dot\beta]$, giving
$SO(8)\times SO(8)$ spinor/co-spinor indices. The distance~$\Delta$ in
128-dimensional Euclidean space of the actual numerically found
stationary point and the position given by the coefficients listed is
also indicated. Furthermore, mass matrix eigenvalues are listed (as
in~\cite{Fischbacher:2002fx}), with multiplicities given in
parentheses where they are greater than~$1$.

The data given contain detailed numerical mass matrix information
about those seven stationary points that were known previously, plus
many new ones, but do not claim to be exhaustive. Also, it is
conceivable (although improbable) that some numerically obtained
stationary point turns out not to correspond to a true exact solution,
but only come close to satisfying the stationarity
constraint. Therefore, the data presented here may be useful as a
first step towards establishing the existence of an analytic
expression for the location and properties of a particular conjectured
vacuum. In principle, the code used here could be modified with
reasonable effort to do computations with high-precision floatingpoint
numbers (\texttt{long double}, or even the GNU Multiprecision
Arithmetics library~\cite{Granlund2004}). As the exponentials of the
spinor coefficients occurring in all analytically known vacua seem to
be algebraic numbers, and as there are simple automatized methods to
guess low-rank polynomials with integer coefficients given a root
known to very high numerical accuracy, going from the numerical data
given here to analytic expressions can be made a semi-automatic
process.\footnote{The code provided with the \LaTeX{} source of the
arXiv preprint of this work contains a simplistic LISP function that
fulfils precisely this purpose. While this function manages to
successfully find the polynomial underlying the algebraic expression
of the spinor coordinates for the $G_2\times G_2$ vacuum, ordinary
IEEE754 double-float numerical accuracy does not contain enough
information to determine with high probability the polynomials
underlying more complicated solutions. This hence would require
upgrading existing code to work with multiprecision floatingpoint
arithmetics.}  It must be emphasized that a numerical approach seems
to be the only feasible way to learn about the properties of those
vacua without any residual gauge symmetry. One may speculate that
these vacua should still possess nontrivial \emph{discrete}
symmetries, which may also allow interesting interpolating solutions.

Due to the choice to perform all computations in double-precision
floatingpoint, rather limited computational effort invested in the
determination of each stationary point, and error accumulation, some
of the numerical data presented in the following tables may be
slightly inaccurate. The most important limitation presumably is that
the given dimension of the residual unbroken gauge group~$H$ can only
be trusted for stationary points `close to the origin': for solutions
far out (with large negative cosmological constant $\Lambda$), this
should only be considered a lower bound on the true dimension. In
particular, the tables show some entries where multiplicities of
eigenvalues of the spinorial mass matrices $A_1$ and $A_3$ would
suggest a nontrivial residual gauge group while ${\rm dim}(H)$ is
listed at zero. For many (but not all) vacua, the eigenvalues of~$A_1$
should be symmetric around zero, which may serve as a first nontrivial
check on the quality of the numerical data. One should remember that
\emph{even if the data for any given vacuum should be slightly off}, 
there is a very high chance that a true stationary point is very close
to the point specified, whose properties may be determined to high
accuracy by restarting the search from the position given with
increased numerical precision.

It should be pointed out that, while extensive searches produced most
of the stationary points with~$R<2.5$ many times over, the solution
with~$SU(4)$ symmetry at~$R\approx2.167$ which already was known
before had to be added manually to the search space: this stationary
point seems to be somewhat non-attractive in a random search. One may
speculate that this might be due to a contribution in the
$(\alpha,\beta)$-spinorial components proportional to the identity,
which could be difficult to produce in a random search. Hence it is
conceivable that the approach used here systematically misses a number
of solutions with large residual symmetry groups that could be
obtained by modifying the way the starting point for the search is
chosen.

Previously known data about the vacuum structure of the model studied
here roughly corresponds to the first page of this lengthy list. All
the other data are novel. Some new features exhibited by these
stationary points that were not encountered so far include total
breaking of all gauge symmetry, as well as $SO(8)\times
SO(8)$-\emph{asymmetric} breaking of supersymmetry. Considering the
sheer amount of new data, the question arises how many different
stationary points there are in this supergravity potential. Judging
from the number of duplicates produced by numerically `fishing for
solutions', the total number that can be found with these methods
presumably is large, but not mind-boggingly so. In particular, given
this list, it already becomes somewhat difficult (though not
impossible) to find further new solutions with~$R<2.5$. The author's
present estimate is that he certainly would accept a $10{:}1$ bet that
the total number is considerably smaller than 10~million. Quite
likely, it is even considerably smaller than~$10\,000$. In particular,
the total number of solutions with some amount of residual
supersymmetry seems to be \emph{very small}, presumably there are
fewer than~25 in total.

\newpage



\begin{longtable}{|l|}\hline
$R=0.000\quad \Lambda=-16.000\quad {\rm SUSY}=(8,8)\quad {\rm dim}(H)=56$\\\hline
\hbox to3.5em{$\Psi_{128}$} \begin{minipage}{24em}{\tiny\vskip1ex{0; $\Delta=0.000000$}\vskip1ex}\end{minipage}\\\hline
\hbox to3.5em{$\mathcal{M}_s/g^2$} \begin{minipage}{24em}{\tiny\vskip1ex{$ -12.000{(\times 128)}$}\vskip1ex}\end{minipage}\\\hline
\hbox to3.5em{$\mathcal{M}_v/g$} \begin{minipage}{24em}{\tiny\vskip1ex{$   0.000{(\times 128)}$}\vskip1ex}\end{minipage}\\\hline
\hbox to3.5em{$A_1$} \begin{minipage}{24em}{\tiny\vskip1ex{$  -2.000{(\times 8)}$, $   2.000{(\times 8)}$}\vskip1ex}\end{minipage}\\\hline
\hbox to3.5em{$A_3$} \begin{minipage}{24em}{\tiny\vskip1ex{$   0.000{(\times 128)}$}\vskip1ex}\end{minipage}\\\hline
\end{longtable}
\smallbreak
\begin{longtable}{|l|}\hline
$R=1.317\quad \Lambda=-25.000\quad {\rm SUSY}=(0,0)\quad {\rm dim}(H)=42$\\\hline
\hbox to3.5em{$\Psi_{128}$} \begin{minipage}{24em}{\tiny\vskip1ex{$1.317[\dot3\dot4]$; $\Delta=0.000042$}\vskip1ex}\end{minipage}\\\hline
\hbox to3.5em{$\mathcal{M}_s/g^2$} \begin{minipage}{24em}{\tiny\vskip1ex{$ -24.000{(\times 49)}$, $  -9.000{(\times 64)}$, $   0.000{(\times 14)}$, $  96.000$}\vskip1ex}\end{minipage}\\\hline
\hbox to3.5em{$\mathcal{M}_v/g$} \begin{minipage}{24em}{\tiny\vskip1ex{$  -6.000{(\times 7)}$, $   0.000{(\times 114)}$, $   6.000{(\times 7)}$}\vskip1ex}\end{minipage}\\\hline
\hbox to3.5em{$A_1$} \begin{minipage}{24em}{\tiny\vskip1ex{$  -3.500{(\times 8)}$, $   3.500{(\times 8)}$}\vskip1ex}\end{minipage}\\\hline
\hbox to3.5em{$A_3$} \begin{minipage}{24em}{\tiny\vskip1ex{$ -10.500{(\times 8)}$, $  -1.500{(\times 56)}$, $   1.500{(\times 56)}$, $  10.500{(\times 8)}$}\vskip1ex}\end{minipage}\\\hline
\end{longtable}
\smallbreak
\begin{longtable}{|l|}\hline
$R=1.491\quad \Lambda=-28.444\quad {\rm SUSY}=(1,1)\quad {\rm dim}(H)=28$\\\hline
\hbox to3.5em{$\Psi_{128}$} \begin{minipage}{24em}{\tiny\vskip1ex{$1.054[83]$; $\Delta=0.000416$}\vskip1ex}\end{minipage}\\\hline
\hbox to3.5em{$\mathcal{M}_s/g^2$} \begin{minipage}{24em}{\tiny\vskip1ex{$ -26.667{(\times 49)}$, $ -12.444{(\times 49)}$, $   0.000{(\times 28)}$, $  16.000$, $ 115.556$}\vskip1ex}\end{minipage}\\\hline
\hbox to3.5em{$\mathcal{M}_v/g$} \begin{minipage}{24em}{\tiny\vskip1ex{$  -6.667{(\times 7)}$, $  -1.333{(\times 7)}$, $   0.000{(\times 100)}$, $   1.333{(\times 7)}$, $   6.667{(\times 7)}$}\vskip1ex}\end{minipage}\\\hline
\hbox to3.5em{$A_1$} \begin{minipage}{24em}{\tiny\vskip1ex{$  -4.000{(\times 7)}$, $  -2.667$, $   2.667$, $   4.000{(\times 7)}$}\vskip1ex}\end{minipage}\\\hline
\hbox to3.5em{$A_3$} \begin{minipage}{24em}{\tiny\vskip1ex{$ -12.000{(\times 7)}$, $  -9.333$, $  -4.000{(\times 7)}$, $  -1.333{(\times 49)}$, $   1.333{(\times 49)}$, $   4.000{(\times 7)}$, $   9.333$, $  12.000{(\times 7)}$}\vskip1ex}\end{minipage}\\\hline
\end{longtable}
\smallbreak
\begin{longtable}{|l|}\hline
$R=1.763\quad \Lambda=-36.000\quad {\rm SUSY}=(2,2)\quad {\rm dim}(H)=18$\\\hline
\hbox to3.5em{$\Psi_{128}$} \begin{minipage}{24em}{\tiny\vskip1ex{$0.881[24]$, $0.881[72]$; $\Delta=0.000748$}\vskip1ex}\end{minipage}\\\hline
\hbox to3.5em{$\mathcal{M}_s/g^2$} \begin{minipage}{24em}{\tiny\vskip1ex{$ -32.000{(\times 49)}$, $ -20.000{(\times 36)}$, $   0.000{(\times 38)}$, $  28.000{(\times 4)}$, $ 160.001$}\vskip1ex}\end{minipage}\\\hline
\hbox to3.5em{$\mathcal{M}_v/g$} \begin{minipage}{24em}{\tiny\vskip1ex{$  -8.000{(\times 7)}$, $  -2.000{(\times 12)}$, $   0.000{(\times 90)}$, $   2.000{(\times 12)}$, $   8.000{(\times 7)}$}\vskip1ex}\end{minipage}\\\hline
\hbox to3.5em{$A_1$} \begin{minipage}{24em}{\tiny\vskip1ex{$  -5.000{(\times 6)}$, $  -3.000{(\times 2)}$, $   3.000{(\times 2)}$, $   5.000{(\times 6)}$}\vskip1ex}\end{minipage}\\\hline
\hbox to3.5em{$A_3$} \begin{minipage}{24em}{\tiny\vskip1ex{$ -15.000{(\times 6)}$, $ -11.000{(\times 2)}$, $  -5.000{(\times 14)}$, $  -1.000{(\times 42)}$, $   1.000{(\times 42)}$, $   5.000{(\times 14)}$, $  11.000{(\times 2)}$, $  15.000{(\times 6)}$}\vskip1ex}\end{minipage}\\\hline
\end{longtable}
\smallbreak
\begin{longtable}{|l|}\hline
$R=2.030\quad \Lambda=-45.753\quad {\rm SUSY}=(0,0)\quad {\rm dim}(H)=8$\\\hline
\hbox to3.5em{$\Psi_{128}$} \begin{minipage}{24em}{\tiny\vskip1ex{$0.459[66]$, $0.014[\dot2\dot3]$, $1.158[\dot3\dot1]$, $0.014[\dot4\dot2]$; $\Delta=0.000443$}\vskip1ex}\end{minipage}\\\hline
\hbox to3.5em{$\mathcal{M}_s/g^2$} \begin{minipage}{24em}{\tiny\vskip1ex{$ -53.352{(\times 5)}$, $ -48.309{(\times 4)}$, $ -47.454{(\times 8)}$, $ -38.084{(\times 16)}$, $ -31.009{(\times 8)}$, $ -27.231{(\times 2)}$, $ -23.007{(\times 16)}$, $ -18.154{(\times 12)}$, $  -2.393{(\times 2)}$, $   0.000{(\times 49)}$, $  83.506$, $  87.608{(\times 2)}$, $ 145.337{(\times 2)}$, $ 197.167$}\vskip1ex}\end{minipage}\\\hline
\hbox to3.5em{$\mathcal{M}_v/g$} \begin{minipage}{24em}{\tiny\vskip1ex{$ -10.661{(\times 3)}$, $  -8.308{(\times 8)}$, $  -4.159{(\times 2)}$, $  -0.925{(\times 2)}$, $  -0.769{(\times 8)}$, $   0.000{(\times 82)}$, $   0.769{(\times 8)}$, $   0.925{(\times 2)}$, $   4.159{(\times 2)}$, $   8.308{(\times 8)}$, $  10.661{(\times 3)}$}\vskip1ex}\end{minipage}\\\hline
\hbox to3.5em{$A_1$} \begin{minipage}{24em}{\tiny\vskip1ex{$  -6.539{(\times 4)}$, $  -4.900{(\times 4)}$, $   4.900{(\times 4)}$, $   6.539{(\times 4)}$}\vskip1ex}\end{minipage}\\\hline
\hbox to3.5em{$A_3$} \begin{minipage}{24em}{\tiny\vskip1ex{$ -19.616{(\times 4)}$, $ -14.699{(\times 4)}$, $ -12.631{(\times 4)}$, $  -8.698{(\times 4)}$, $  -5.463{(\times 4)}$, $  -4.539{(\times 4)}$, $  -3.614{(\times 4)}$, $  -3.000{(\times 8)}$, $  -2.900{(\times 4)}$, $  -1.361{(\times 16)}$, $  -0.831{(\times 4)}$, $  -0.379{(\times 4)}$, $   0.379{(\times 4)}$, $   0.831{(\times 4)}$, $   1.361{(\times 16)}$, $   2.900{(\times 4)}$, $   3.000{(\times 8)}$, $   3.614{(\times 4)}$, $   4.539{(\times 4)}$, $   5.463{(\times 4)}$, $   8.698{(\times 4)}$, $  12.631{(\times 4)}$, $  14.699{(\times 4)}$, $  19.616{(\times 4)}$}\vskip1ex}\end{minipage}\\\hline
\end{longtable}
\smallbreak
\begin{longtable}{|l|}\hline
$R=2.122\quad \Lambda=-49.821\quad {\rm SUSY}=(0,0)\quad {\rm dim}(H)=10$\\\hline
\hbox to3.5em{$\Psi_{128}$} \begin{minipage}{24em}{\tiny\vskip1ex{$0.462[22]$, $0.462[63]$, $0.462[86]$; $\Delta=0.001035$}\vskip1ex}\end{minipage}\\\hline
\hbox to3.5em{$\mathcal{M}_s/g^2$} \begin{minipage}{24em}{\tiny\vskip1ex{$ -65.923$, $ -56.599{(\times 12)}$, $ -46.978{(\times 16)}$, $ -39.397{(\times 6)}$, $ -28.652{(\times 6)}$, $ -24.861{(\times 2)}$, $ -20.365{(\times 12)}$, $ -17.489{(\times 4)}$, $  -6.248{(\times 12)}$, $   0.000{(\times 46)}$, $  61.095{(\times 8)}$, $ 210.857$, $ 220.146{(\times 2)}$}\vskip1ex}\end{minipage}\\\hline
\hbox to3.5em{$\mathcal{M}_v/g$} \begin{minipage}{24em}{\tiny\vskip1ex{$ -11.557$, $ -10.287{(\times 6)}$, $  -8.287{(\times 6)}$, $  -0.457{(\times 6)}$, $   0.000{(\times 90)}$, $   0.457{(\times 6)}$, $   8.287{(\times 6)}$, $  10.287{(\times 6)}$, $  11.557$}\vskip1ex}\end{minipage}\\\hline
\hbox to3.5em{$A_1$} \begin{minipage}{24em}{\tiny\vskip1ex{$  -7.215{(\times 2)}$, $  -5.987{(\times 6)}$, $   5.987{(\times 6)}$, $   7.215{(\times 2)}$}\vskip1ex}\end{minipage}\\\hline
\hbox to3.5em{$A_3$} \begin{minipage}{24em}{\tiny\vskip1ex{$ -21.646{(\times 2)}$, $ -17.960{(\times 6)}$, $ -13.359{(\times 6)}$, $ -10.588{(\times 2)}$, $  -4.758{(\times 12)}$, $  -3.843{(\times 2)}$, $  -3.529{(\times 16)}$, $  -2.614{(\times 6)}$, $  -1.386{(\times 6)}$, $  -1.072{(\times 6)}$, $   1.072{(\times 6)}$, $   1.386{(\times 6)}$, $   2.614{(\times 6)}$, $   3.529{(\times 16)}$, $   3.843{(\times 2)}$, $   4.758{(\times 12)}$, $  10.588{(\times 2)}$, $  13.359{(\times 6)}$, $  17.960{(\times 6)}$, $  21.646{(\times 2)}$}\vskip1ex}\end{minipage}\\\hline
\end{longtable}
\smallbreak
\begin{longtable}{|l|}\hline
$R=2.167\quad \Lambda=-52.000\quad {\rm SUSY}=(0,0)\quad {\rm dim}(H)=15$\\\hline
\hbox to3.5em{$\Psi_{128}$} \begin{minipage}{24em}{\tiny\vskip1ex{$0.392[11]$, $0.392[22]$, $0.392[33]$, $0.392[44]$, $0.392[55]$, $0.392[66]$, $0.392[77]$, $0.392[88]$, $1.317[\dot4\dot6]$, $1.317[\dot6\dot4]$; $\Delta=0.000851$}\vskip1ex}\end{minipage}\\\hline
\hbox to3.5em{$\mathcal{M}_s/g^2$} \begin{minipage}{24em}{\tiny\vskip1ex{$ -73.986$, $ -60.000{(\times 20)}$, $ -36.000{(\times 15)}$, $ -33.000{(\times 12)}$, $ -13.986$, $  -9.000{(\times 20)}$, $   0.000{(\times 41)}$, $  48.000{(\times 15)}$, $ 217.986$, $ 252.000$, $ 277.986$}\vskip1ex}\end{minipage}\\\hline
\hbox to3.5em{$\mathcal{M}_v/g$} \begin{minipage}{24em}{\tiny\vskip1ex{$ -12.000$, $  -9.950{(\times 12)}$, $   0.000{(\times 102)}$, $   9.950{(\times 12)}$, $  12.000$}\vskip1ex}\end{minipage}\\\hline
\hbox to3.5em{$A_1$} \begin{minipage}{24em}{\tiny\vskip1ex{$  -6.614{(\times 8)}$, $   6.614{(\times 8)}$}\vskip1ex}\end{minipage}\\\hline
\hbox to3.5em{$A_3$} \begin{minipage}{24em}{\tiny\vskip1ex{$ -19.843{(\times 8)}$, $ -13.747{(\times 8)}$, $  -3.969{(\times 40)}$, $  -2.128{(\times 8)}$, $   2.128{(\times 8)}$, $   3.969{(\times 40)}$, $  13.747{(\times 8)}$, $  19.843{(\times 8)}$}\vskip1ex}\end{minipage}\\\hline
\end{longtable}
\smallbreak
\begin{longtable}{|l|}\hline
$R=2.281\quad \Lambda=-60.063\quad {\rm SUSY}=(0,1)\quad {\rm dim}(H)=6$\\\hline
\hbox to3.5em{$\Psi_{128}$} \begin{minipage}{24em}{\tiny\vskip1ex{$0.899[55]$, $0.899[78]$, $0.899[\dot1\dot7]$, $0.899[\dot3\dot6]$, $0.931[\dot8\dot2]$; $\Delta=0.001051$}\vskip1ex}\end{minipage}\\\hline
\hbox to3.5em{$\mathcal{M}_s/g^2$} \begin{minipage}{24em}{\tiny\vskip1ex{$ -60.000{(\times 5)}$, $ -59.500$, $ -55.000{(\times 3)}$, $ -52.500{(\times 24)}$, $ -40.338{(\times 5)}$, $ -37.500{(\times 8)}$, $ -27.000{(\times 16)}$, $ -25.127{(\times 5)}$, $   0.000{(\times 50)}$, $  70.838{(\times 5)}$, $ 167.627{(\times 5)}$, $ 273.000$}\vskip1ex}\end{minipage}\\\hline
\hbox to3.5em{$\mathcal{M}_v/g$} \begin{minipage}{24em}{\tiny\vskip1ex{$ -13.500{(\times 3)}$, $ -10.000{(\times 8)}$, $  -4.500{(\times 4)}$, $  -3.000{(\times 3)}$, $  -2.000{(\times 3)}$, $  -1.000{(\times 4)}$, $   0.000{(\times 78)}$, $   1.000{(\times 7)}$, $   4.500{(\times 7)}$, $  10.000{(\times 5)}$, $  11.000{(\times 3)}$, $  12.000{(\times 3)}$}\vskip1ex}\end{minipage}\\\hline
\hbox to3.5em{$A_1$} \begin{minipage}{24em}{\tiny\vskip1ex{$  -8.375{(\times 4)}$, $  -5.875{(\times 3)}$, $  -3.875$, $   6.125{(\times 5)}$, $   8.125{(\times 3)}$}\vskip1ex}\end{minipage}\\\hline
\hbox to3.5em{$A_3$} \begin{minipage}{24em}{\tiny\vskip1ex{$ -25.125{(\times 4)}$, $ -17.625{(\times 3)}$, $ -15.316{(\times 5)}$, $  -9.625{(\times 3)}$, $  -8.625{(\times 8)}$, $  -6.125{(\times 8)}$, $  -3.625{(\times 5)}$, $  -3.125$, $  -1.625{(\times 3)}$, $  -1.125{(\times 24)}$, $   0.566{(\times 5)}$, $   0.875{(\times 3)}$, $   1.875{(\times 16)}$, $   2.036{(\times 5)}$, $   2.875{(\times 4)}$, $   4.875{(\times 7)}$, $   8.375{(\times 7)}$, $  11.214{(\times 5)}$, $  14.375$, $  14.875{(\times 3)}$, $  18.375{(\times 5)}$, $  24.375{(\times 3)}$}\vskip1ex}\end{minipage}\\\hline
\end{longtable}
\smallbreak
\begin{longtable}{|l|}\hline
$R=2.339\quad \Lambda=-66.459\quad {\rm SUSY}=(0,0)\quad {\rm dim}(H)=9$\\\hline
\hbox to3.5em{$\Psi_{128}$} \begin{minipage}{24em}{\tiny\vskip1ex{$0.136[13]$, $1.071[22]$, $0.744[71]$, $0.698[\dot3\dot1]$; $\Delta=0.001254$}\vskip1ex}\end{minipage}\\\hline
\hbox to3.5em{$\mathcal{M}_s/g^2$} \begin{minipage}{24em}{\tiny\vskip1ex{$ -76.717{(\times 5)}$, $ -68.398{(\times 4)}$, $ -53.031{(\times 12)}$, $ -52.997{(\times 8)}$, $ -48.185{(\times 16)}$, $ -47.007$, $ -41.121{(\times 4)}$, $ -39.320{(\times 12)}$, $ -13.593{(\times 8)}$, $   0.000{(\times 47)}$, $  50.092{(\times 5)}$, $  83.699$, $ 167.796{(\times 3)}$, $ 202.491$, $ 339.258$}\vskip1ex}\end{minipage}\\\hline
\hbox to3.5em{$\mathcal{M}_v/g$} \begin{minipage}{24em}{\tiny\vskip1ex{$ -13.158$, $ -13.157{(\times 2)}$, $ -12.026{(\times 4)}$, $  -8.483{(\times 4)}$, $  -4.346{(\times 3)}$, $  -2.801{(\times 8)}$, $   0.000{(\times 84)}$, $   2.801{(\times 8)}$, $   4.346{(\times 3)}$, $   8.483{(\times 4)}$, $  12.026{(\times 4)}$, $  13.157{(\times 2)}$, $  13.158$}\vskip1ex}\end{minipage}\\\hline
\hbox to3.5em{$A_1$} \begin{minipage}{24em}{\tiny\vskip1ex{$  -8.528{(\times 4)}$, $  -7.867$, $  -4.642{(\times 3)}$, $   4.642{(\times 3)}$, $   7.867$, $   8.528{(\times 4)}$}\vskip1ex}\end{minipage}\\
\hline
\end{longtable}
\smallbreak
\begin{longtable}{|l|}\hline
$R=2.370\quad \Lambda=-65.744\quad {\rm SUSY}=(0,0)\quad {\rm dim}(H)=1$\\\hline
\hbox to3.5em{$\Psi_{128}$} \begin{minipage}{24em}{\tiny\vskip1ex{$0.003[17]$, $0.100[21]$, $0.295[22]$, $0.005[23]$, $0.955[26]$, $0.003[28]$, $0.001[31]$, $0.021[33]$, $0.001[36]$, $0.004[42]$, $0.265[45]$, $0.009[51]$, $0.001[54]$, $0.003[61]$, $0.005[66]$, $0.041[71]$, $0.121[72]$, $0.002[73]$, $0.123[76]$, $0.001[78]$, $0.001[85]$, $0.004[88]$, $0.001[\dot1\dot2]$, $0.001[\dot1\dot4]$, $0.027[\dot1\dot7]$, $0.001[\dot2\dot3]$, $0.002[\dot2\dot8]$, $0.001[\dot3\dot1]$, $0.026[\dot3\dot3]$, $0.076[\dot3\dot4]$, $0.001[\dot3\dot6]$, $0.013[\dot4\dot6]$, $0.009[\dot5\dot6]$, $0.004[\dot6\dot5]$, $0.003[\dot7\dot3]$, $0.005[\dot7\dot8]$, $0.009[\dot8\dot3]$, $0.014[\dot8\dot7]$; $\Delta=0.002569$}\vskip1ex}\end{minipage}\\\hline
\hbox to3.5em{$\mathcal{M}_s/g^2$} \begin{minipage}{24em}{\tiny\vskip1ex{$ -78.457$, $ -77.063$, $ -75.306$, $ -69.065$, $ -65.952{(\times 4)}$, $ -62.258{(\times 4)}$, $ -60.856$, $ -59.842{(\times 2)}$, $ -57.429$, $ -57.035{(\times 2)}$, $ -55.290$, $ -54.648{(\times 4)}$, $ -53.500{(\times 2)}$, $ -52.928{(\times 2)}$, $ -51.380{(\times 2)}$, $ -49.712{(\times 2)}$, $ -44.922{(\times 4)}$, $ -39.781{(\times 2)}$, $ -37.424{(\times 2)}$, $ -36.647$, $ -36.239{(\times 4)}$, $ -29.970{(\times 4)}$, $ -26.558$, $ -26.364{(\times 4)}$, $ -25.874$, $ -25.566{(\times 2)}$, $ -24.244{(\times 2)}$, $ -24.133$, $ -22.533$, $ -18.808$, $   0.000{(\times 55)}$, $  70.708$, $  73.569$, $ 133.551$, $ 142.236{(\times 4)}$, $ 151.200{(\times 2)}$, $ 252.334{(\times 2)}$, $ 293.903$}\vskip1ex}\end{minipage}\\
\hline
\end{longtable}
\smallbreak
\begin{longtable}{|l|}\hline
$R=2.411\quad \Lambda=-71.204\quad {\rm SUSY}=(1,0)\quad {\rm dim}(H)=5$\\\hline
\hbox to3.5em{$\Psi_{128}$} \begin{minipage}{24em}{\tiny\vskip1ex{$1.009[17]$, $0.611[24]$, $1.009[81]$, $0.083[\dot3\dot1]$; $\Delta=0.000955$}\vskip1ex}\end{minipage}\\\hline
\hbox to3.5em{$\mathcal{M}_s/g^2$} \begin{minipage}{24em}{\tiny\vskip1ex{$ -71.151{(\times 4)}$, $ -70.963{(\times 2)}$, $ -70.395{(\times 5)}$, $ -66.463{(\times 4)}$, $ -61.439{(\times 4)}$, $ -59.453{(\times 3)}$, $ -58.086{(\times 4)}$, $ -56.997{(\times 2)}$, $ -54.963{(\times 8)}$, $ -48.776{(\times 2)}$, $ -47.198{(\times 4)}$, $ -47.197{(\times 4)}$, $ -37.920{(\times 8)}$, $ -32.855{(\times 2)}$, $ -26.802{(\times 4)}$, $ -17.799{(\times 4)}$, $   0.000{(\times 51)}$, $  34.698{(\times 2)}$, $  66.419{(\times 2)}$, $ 103.761{(\times 3)}$, $ 114.278$, $ 179.367{(\times 2)}$, $ 241.354{(\times 2)}$, $ 352.963$}\vskip1ex}\end{minipage}\\\hline
\hbox to3.5em{$\mathcal{M}_v/g$} \begin{minipage}{24em}{\tiny\vskip1ex{$ -13.338{(\times 3)}$, $ -12.930{(\times 2)}$, $ -12.208{(\times 2)}$, $ -11.077$, $  -9.799{(\times 3)}$, $  -6.669{(\times 2)}$, $  -5.130{(\times 2)}$, $  -3.010{(\times 2)}$, $  -2.261{(\times 5)}$, $  -1.393{(\times 3)}$, $  -0.408{(\times 2)}$, $   0.000{(\times 77)}$, $   1.130{(\times 2)}$, $   2.261{(\times 4)}$, $   3.539$, $   4.549{(\times 2)}$, $   5.130{(\times 4)}$, $   9.799{(\times 4)}$, $  12.208{(\times 4)}$, $  14.731{(\times 3)}$}\vskip1ex}\end{minipage}\\\hline
\hbox to3.5em{$A_1$} \begin{minipage}{24em}{\tiny\vskip1ex{$  -9.119{(\times 3)}$, $  -7.988{(\times 2)}$, $  -5.580{(\times 3)}$, $   4.219$, $   5.350{(\times 2)}$, $   7.758$, $   9.350{(\times 4)}$}\vskip1ex}\end{minipage}\\
\hline
\end{longtable}
\smallbreak
\begin{longtable}{|l|}\hline
$R=2.434\quad \Lambda=-70.269\quad {\rm SUSY}=(0,0)\quad {\rm dim}(H)=2$\\\hline
\hbox to3.5em{$\Psi_{128}$} \begin{minipage}{24em}{\tiny\vskip1ex{$0.093[38]$, $0.005[55]$, $0.093[82]$, $0.945[\dot3\dot7]$, $0.005[\dot5\dot6]$, $0.093[\dot7\dot8]$; $\Delta=0.000638$}\vskip1ex}\end{minipage}\\\hline
\hbox to3.5em{$\mathcal{M}_s/g^2$} \begin{minipage}{24em}{\tiny\vskip1ex{$ -91.259$, $ -82.458{(\times 4)}$, $ -65.898{(\times 4)}$, $ -63.223{(\times 4)}$, $ -62.305$, $ -60.951{(\times 4)}$, $ -59.727$, $ -57.647{(\times 8)}$, $ -47.952{(\times 4)}$, $ -47.419{(\times 4)}$, $ -38.654{(\times 4)}$, $ -35.631{(\times 2)}$, $ -34.282{(\times 4)}$, $ -27.232$, $ -27.231{(\times 3)}$, $ -23.730{(\times 4)}$, $ -23.308$, $ -23.049{(\times 4)}$, $ -13.366$, $   0.000{(\times 54)}$, $  63.933{(\times 4)}$, $ 120.950{(\times 4)}$, $ 174.272$, $ 220.893{(\times 4)}$, $ 288.134$, $ 312.009$}\vskip1ex}\end{minipage}\\\hline
\hbox to3.5em{$\mathcal{M}_v/g$} \begin{minipage}{24em}{\tiny\vskip1ex{$ -13.653{(\times 4)}$, $ -13.546$, $ -12.289{(\times 2)}$, $ -11.352{(\times 2)}$, $ -10.668$, $  -9.515{(\times 4)}$, $  -5.714{(\times 2)}$, $  -2.698{(\times 4)}$, $  -1.156{(\times 2)}$, $  -0.742$, $  -0.646{(\times 4)}$, $   0.000{(\times 74)}$, $   0.646{(\times 4)}$, $   0.742$, $   1.156{(\times 2)}$, $   2.698{(\times 4)}$, $   5.714{(\times 2)}$, $   9.515{(\times 4)}$, $  10.668$, $  11.352{(\times 2)}$, $  12.289{(\times 2)}$, $  13.546$, $  13.653{(\times 4)}$}\vskip1ex}\end{minipage}\\\hline
\hbox to3.5em{$A_1$} \begin{minipage}{24em}{\tiny\vskip1ex{$  -9.941$, $  -8.339{(\times 4)}$, $  -7.439$, $  -6.650{(\times 2)}$, $   6.650{(\times 2)}$, $   7.439$, $   8.339{(\times 4)}$, $   9.941$}\vskip1ex}\end{minipage}\\
\hline
\end{longtable}
\smallbreak
\begin{longtable}{|l|}\hline
$R=2.440\quad \Lambda=-72.983\quad {\rm SUSY}=(0,0)\quad {\rm dim}(H)=2$\\\hline
\hbox to3.5em{$\Psi_{128}$} \begin{minipage}{24em}{\tiny\vskip1ex{$0.001[26]$, $0.615[27]$, $0.001[32]$, $0.001[48]$, $0.001[81]$, $0.309[\dot1\dot8]$, $0.178[\dot2\dot2]$, $0.307[\dot3\dot7]$, $0.001[\dot5\dot4]$, $0.309[\dot6\dot7]$, $0.001[\dot7\dot2]$, $0.178[\dot7\dot3]$, $0.307[\dot8\dot8]$; $\Delta=0.001933$}\vskip1ex}\end{minipage}\\\hline
\hbox to3.5em{$\mathcal{M}_s/g^2$} \begin{minipage}{24em}{\tiny\vskip1ex{$ -84.872{(\times 2)}$, $ -83.516$, $ -75.878{(\times 4)}$, $ -72.056{(\times 2)}$, $ -72.055{(\times 2)}$, $ -69.407$, $ -69.311{(\times 2)}$, $ -67.582{(\times 2)}$, $ -64.251{(\times 4)}$, $ -63.978$, $ -60.301{(\times 4)}$, $ -59.912$, $ -58.070{(\times 2)}$, $ -51.152{(\times 2)}$, $ -49.569{(\times 2)}$, $ -46.358{(\times 4)}$, $ -46.088{(\times 2)}$, $ -46.008$, $ -45.683{(\times 3)}$, $ -45.682$, $ -40.375{(\times 4)}$, $ -36.128$, $ -35.718{(\times 2)}$, $ -35.247{(\times 2)}$, $ -29.958{(\times 2)}$, $ -22.583{(\times 4)}$, $ -16.807{(\times 2)}$, $   0.000{(\times 54)}$, $  27.193{(\times 2)}$, $  42.845$, $ 101.245$, $ 104.250$, $ 115.806{(\times 2)}$, $ 159.468{(\times 4)}$, $ 290.616{(\times 2)}$, $ 354.487$}\vskip1ex}\end{minipage}\\
\hline
\end{longtable}
\smallbreak
\begin{longtable}{|l|}\hline
$R=2.491\quad \Lambda=-75.557\quad {\rm SUSY}=(0,0)\quad {\rm dim}(H)=2$\\\hline
\hbox to3.5em{$\Psi_{128}$} \begin{minipage}{24em}{\tiny\vskip1ex{$0.216[14]$, $0.017[15]$, $0.763[25]$, $0.034[32]$, $0.216[38]$, $0.034[43]$, $0.216[47]$, $0.763[56]$, $0.034[65]$, $0.060[67]$, $0.216[76]$, $0.034[81]$, $0.060[83]$, $0.017[84]$, $0.023[\dot1\dot2]$, $0.002[\dot1\dot7]$, $1.079[\dot2\dot3]$, $0.074[\dot3\dot4]$, $0.006[\dot3\dot5]$, $0.003[\dot3\dot8]$, $0.002[\dot4\dot2]$, $0.002[\dot4\dot5]$, $0.023[\dot5\dot2]$, $0.002[\dot5\dot7]$, $0.074[\dot7\dot4]$, $0.006[\dot7\dot5]$, $0.003[\dot7\dot6]$, $0.002[\dot8\dot2]$, $0.002[\dot8\dot5]$; $\Delta=0.002500$}\vskip1ex}\end{minipage}\\\hline
\hbox to3.5em{$\mathcal{M}_s/g^2$} \begin{minipage}{24em}{\tiny\vskip1ex{$ -90.613$, $ -87.470{(\times 2)}$, $ -81.455$, $ -77.977{(\times 2)}$, $ -76.164{(\times 2)}$, $ -75.268$, $ -74.246{(\times 2)}$, $ -70.508{(\times 2)}$, $ -65.907{(\times 2)}$, $ -64.320{(\times 6)}$, $ -62.397{(\times 2)}$, $ -55.538$, $ -54.660{(\times 4)}$, $ -51.442{(\times 2)}$, $ -45.886{(\times 2)}$, $ -45.557$, $ -45.199{(\times 4)}$, $ -43.665{(\times 2)}$, $ -37.078{(\times 2)}$, $ -35.507{(\times 2)}$, $ -35.052{(\times 2)}$, $ -34.746$, $ -32.420{(\times 2)}$, $ -31.788$, $ -31.012{(\times 2)}$, $ -26.032{(\times 2)}$, $ -22.932{(\times 4)}$, $ -21.068{(\times 2)}$, $   0.000{(\times 54)}$, $  56.667{(\times 2)}$, $ 103.172$, $ 103.173$, $ 147.343{(\times 4)}$, $ 147.980$, $ 225.472$, $ 228.232{(\times 2)}$, $ 280.984{(\times 2)}$, $ 340.472$}\vskip1ex}\end{minipage}\\
\hline
\end{longtable}
\smallbreak
\begin{longtable}{|l|}\hline
$R=2.515\quad \Lambda=-77.399\quad {\rm SUSY}=(0,0)\quad {\rm dim}(H)=0$\\\hline
\hbox to3.5em{$\Psi_{128}$} \begin{minipage}{24em}{\tiny\vskip1ex{$0.208[28]$, $0.053[35]$, $0.074[48]$, $0.890[51]$, $0.208[63]$, $0.704[64]$, $0.890[72]$, $0.074[83]$, $0.208[84]$, $0.006[\dot2\dot1]$, $0.049[\dot2\dot2]$, $1.113[\dot4\dot2]$, $0.049[\dot6\dot5]$, $0.075[\dot7\dot8]$, $0.006[\dot8\dot5]$, $0.049[\dot8\dot6]$; $\Delta=0.001207$}\vskip1ex}\end{minipage}\\\hline
\hbox to3.5em{$\mathcal{M}_s/g^2$} \begin{minipage}{24em}{\tiny\vskip1ex{$ -94.774$, $ -89.755$, $ -89.754$, $ -88.089$, $ -87.621$, $ -87.465$, $ -77.255$, $ -76.974{(\times 2)}$, $ -76.209{(\times 2)}$, $ -73.198$, $ -71.585$, $ -70.913$, $ -70.552{(\times 2)}$, $ -67.038$, $ -64.048$, $ -62.000$, $ -61.905$, $ -60.495{(\times 2)}$, $ -58.376$, $ -58.344{(\times 2)}$, $ -56.402{(\times 2)}$, $ -54.745$, $ -51.266$, $ -51.014$, $ -49.118$, $ -48.984{(\times 2)}$, $ -47.771$, $ -43.489$, $ -43.222$, $ -42.864$, $ -41.170$, $ -40.555{(\times 2)}$, $ -36.529$, $ -36.528$, $ -34.844{(\times 2)}$, $ -30.154{(\times 2)}$, $ -29.770$, $ -29.045$, $ -28.964$, $ -28.216$, $ -27.953$, $ -26.301$, $ -23.071{(\times 2)}$, $ -19.752$, $ -18.763$, $   0.000{(\times 56)}$, $  42.474$, $  58.267$, $ 105.663{(\times 2)}$, $ 110.680$, $ 132.824$, $ 137.345$, $ 145.530{(\times 2)}$, $ 198.922$, $ 204.536$, $ 245.812$, $ 245.813$, $ 283.108$, $ 324.781$, $ 346.046$}\vskip1ex}\end{minipage}\\
\hline
\end{longtable}
\smallbreak
\begin{longtable}{|l|}\hline
$R=2.534\quad \Lambda=-78.875\quad {\rm SUSY}=(0,0)\quad {\rm dim}(H)=2$\\\hline
\hbox to3.5em{$\Psi_{128}$} \begin{minipage}{24em}{\tiny\vskip1ex{$0.825[12]$, $-0.069[26]$, $0.069[37]$, $-0.069[43]$, $0.825[58]$, $-0.825[64]$, $-0.825[71]$, $0.069[85]$, $-0.005[\dot1\dot6]$, $-1.153[\dot2\dot2]$, $0.005[\dot3\dot3]$, $1.008[\dot4\dot7]$, $-0.082[\dot5\dot4]$, $0.005[\dot6\dot8]$, $1.153[\dot7\dot5]$, $0.005[\dot8\dot1]$; $\Delta=0.000714$}\vskip1ex}\end{minipage}\\\hline
\hbox to3.5em{$\mathcal{M}_s/g^2$} \begin{minipage}{24em}{\tiny\vskip1ex{$-101.966$, $ -95.668{(\times 4)}$, $ -85.332$, $ -83.178{(\times 4)}$, $ -77.936{(\times 2)}$, $ -71.825{(\times 4)}$, $ -70.814$, $ -61.307{(\times 4)}$, $ -57.793$, $ -56.584$, $ -54.488$, $ -53.356{(\times 4)}$, $ -49.458{(\times 4)}$, $ -46.077{(\times 4)}$, $ -45.509{(\times 2)}$, $ -40.337{(\times 4)}$, $ -33.701{(\times 2)}$, $ -30.127{(\times 4)}$, $ -27.039$, $ -23.613$, $ -19.692{(\times 4)}$, $ -18.615$, $   0.000{(\times 54)}$, $  40.014{(\times 4)}$, $  86.707{(\times 4)}$, $ 137.635$, $ 176.645$, $ 177.956{(\times 4)}$, $ 223.790{(\times 2)}$, $ 338.103$, $ 349.795$, $ 354.699$}\vskip1ex}\end{minipage}\\\hline
\hbox to3.5em{$\mathcal{M}_v/g$} \begin{minipage}{24em}{\tiny\vskip1ex{$ -14.679$, $ -14.511{(\times 2)}$, $ -14.034{(\times 4)}$, $ -11.648{(\times 4)}$, $ -11.492{(\times 2)}$, $ -10.303$, $  -5.446{(\times 4)}$, $  -1.922{(\times 4)}$, $  -1.060{(\times 4)}$, $   0.000{(\times 76)}$, $   1.060{(\times 4)}$, $   1.922{(\times 4)}$, $   5.446{(\times 4)}$, $  10.303$, $  11.492{(\times 2)}$, $  11.648{(\times 4)}$, $  14.034{(\times 4)}$, $  14.511{(\times 2)}$, $  14.679$}\vskip1ex}\end{minipage}\\\hline
\hbox to3.5em{$A_1$} \begin{minipage}{24em}{\tiny\vskip1ex{$  -9.765{(\times 4)}$, $  -7.596{(\times 4)}$, $   7.596{(\times 4)}$, $   9.765{(\times 4)}$}\vskip1ex}\end{minipage}\\
\hline
\end{longtable}
\smallbreak
\begin{longtable}{|l|}\hline
$R=2.546\quad \Lambda=-81.651\quad {\rm SUSY}=(0,0)\quad {\rm dim}(H)=1$\\\hline
\hbox to3.5em{$\Psi_{128}$} \begin{minipage}{24em}{\tiny\vskip1ex{$-0.024[11]$, $0.319[12]$, $-0.000[13]$, $0.217[17]$, $-0.148[18]$, $0.130[21]$, $0.220[22]$, $0.171[27]$, $-0.125[28]$, $-0.058[33]$, $-0.007[34]$, $-0.027[35]$, $-0.016[36]$, $-0.050[43]$, $-0.953[44]$, $0.257[45]$, $0.076[46]$, $0.193[53]$, $0.059[54]$, $-0.021[55]$, $0.971[56]$, $0.000[62]$, $0.036[63]$, $-0.042[64]$, $0.030[65]$, $0.020[66]$, $0.000[67]$, $-0.971[71]$, $0.021[72]$, $-0.000[75]$, $-0.129[77]$, $0.027[78]$, $-0.052[81]$, $0.079[82]$, $0.044[87]$, $-0.115[88]$, $0.074[\dot1\dot1]$, $0.190[\dot1\dot2]$, $0.000[\dot1\dot5]$, $-0.000[\dot1\dot6]$, $0.241[\dot1\dot7]$, $0.486[\dot1\dot8]$, $-0.073[\dot2\dot1]$, $0.050[\dot2\dot2]$, $-0.034[\dot2\dot7]$, $-0.133[\dot2\dot8]$, $-0.004[\dot3\dot3]$, $-0.963[\dot3\dot4]$, $-0.202[\dot3\dot5]$, $-0.005[\dot3\dot6]$, $0.004[\dot4\dot3]$, $-0.017[\dot4\dot4]$, $0.067[\dot4\dot5]$, $-0.021[\dot4\dot6]$, $-0.945[\dot5\dot3]$, $-0.029[\dot5\dot4]$, $0.140[\dot5\dot5]$, $-0.261[\dot5\dot6]$, $-0.000[\dot5\dot7]$, $0.033[\dot6\dot3]$, $-0.133[\dot6\dot4]$, $-0.032[\dot6\dot5]$, $0.008[\dot6\dot6]$, $0.000[\dot6\dot7]$, $0.000[\dot6\dot8]$, $0.004[\dot7\dot1]$, $0.050[\dot7\dot2]$, $-0.101[\dot7\dot7]$, $-0.128[\dot7\dot8]$, $-0.056[\dot8\dot1]$, $0.914[\dot8\dot2]$, $0.039[\dot8\dot7]$, $0.189[\dot8\dot8]$; $\Delta=0.002179$}\vskip1ex}\end{minipage}\\
\hline
\end{longtable}
\smallbreak
\begin{longtable}{|l|}\hline
$R=2.554\quad \Lambda=-82.168\quad {\rm SUSY}=(0,0)\quad {\rm dim}(H)=0$\\\hline
\hbox to3.5em{$\Psi_{128}$} \begin{minipage}{24em}{\tiny\vskip1ex{$0.003[14]$, $0.001[15]$, $0.075[16]$, $1.011[17]$, $0.954[21]$, $0.002[23]$, $0.005[26]$, $0.002[28]$, $0.030[31]$, $0.002[35]$, $0.004[37]$, $0.001[38]$, $0.002[44]$, $0.031[46]$, $0.111[47]$, $0.001[51]$, $0.215[52]$, $0.002[55]$, $0.003[57]$, $0.001[66]$, $0.002[71]$, $0.001[73]$, $0.386[82]$, $0.031[\dot1\dot3]$, $0.005[\dot1\dot5]$, $0.004[\dot2\dot2]$, $0.001[\dot2\dot3]$, $0.030[\dot2\dot5]$, $0.001[\dot2\dot6]$, $0.954[\dot3\dot5]$, $0.001[\dot3\dot6]$, $0.001[\dot4\dot1]$, $0.076[\dot4\dot3]$, $0.002[\dot4\dot5]$, $0.002[\dot5\dot2]$, $0.001[\dot6\dot5]$, $0.006[\dot6\dot6]$, $0.001[\dot6\dot8]$, $0.076[\dot7\dot1]$, $0.031[\dot7\dot4]$, $0.002[\dot7\dot5]$, $0.001[\dot8\dot2]$, $0.001[\dot8\dot4]$, $0.001[\dot8\dot5]$, $0.215[\dot8\dot6]$, $0.387[\dot8\dot7]$; $\Delta=0.003336$}\vskip1ex}\end{minipage}\\\hline
\hbox to3.5em{$\mathcal{M}_s/g^2$} \begin{minipage}{24em}{\tiny\vskip1ex{$ -92.643$, $ -89.925$, $ -87.938{(\times 2)}$, $ -87.535$, $ -86.629{(\times 2)}$, $ -82.550$, $ -80.198$, $ -80.165$, $ -79.645$, $ -78.365$, $ -73.668{(\times 2)}$, $ -71.430{(\times 2)}$, $ -71.406$, $ -71.268$, $ -71.038{(\times 2)}$, $ -69.154$, $ -68.822{(\times 2)}$, $ -67.673$, $ -67.101$, $ -59.890$, $ -58.426{(\times 2)}$, $ -57.479$, $ -56.823$, $ -55.054{(\times 2)}$, $ -52.666$, $ -51.273$, $ -50.177$, $ -48.906{(\times 2)}$, $ -46.893$, $ -45.254$, $ -45.219$, $ -43.000$, $ -41.287{(\times 2)}$, $ -38.060$, $ -37.348{(\times 2)}$, $ -34.427$, $ -33.859{(\times 2)}$, $ -33.777$, $ -29.016$, $ -26.871$, $ -26.870$, $ -26.869$, $ -16.736$, $   0.000{(\times 56)}$, $  21.567$, $  29.888$, $  84.557$, $ 100.326$, $ 114.263$, $ 119.334{(\times 2)}$, $ 125.095$, $ 140.703$, $ 140.704$, $ 185.336$, $ 246.054$, $ 275.277$, $ 275.278$, $ 280.739$, $ 330.856$, $ 384.086$}\vskip1ex}\end{minipage}\\
\hline
\end{longtable}
\smallbreak
\begin{longtable}{|l|}\hline
$R=2.563\quad \Lambda=-81.767\quad {\rm SUSY}=(0,0)\quad {\rm dim}(H)=1$\\\hline
\hbox to3.5em{$\Psi_{128}$} \begin{minipage}{24em}{\tiny\vskip1ex{$0.027[26]$, $0.001[27]$, $0.011[31]$, $0.002[34]$, $0.001[35]$, $0.034[38]$, $0.001[43]$, $0.001[44]$, $1.203[45]$, $0.001[54]$, $0.001[63]$, $0.021[71]$, $0.001[73]$, $0.002[75]$, $0.042[77]$, $0.007[83]$, $0.002[85]$, $0.028[86]$, $0.001[87]$, $0.029[\dot1\dot1]$, $0.024[\dot1\dot2]$, $0.010[\dot1\dot4]$, $0.035[\dot1\dot5]$, $0.015[\dot1\dot6]$, $0.109[\dot2\dot2]$, $0.010[\dot2\dot6]$, $0.057[\dot3\dot1]$, $0.029[\dot3\dot2]$, $0.078[\dot3\dot3]$, $0.081[\dot3\dot5]$, $0.030[\dot3\dot6]$, $0.071[\dot3\dot7]$, $0.053[\dot3\dot8]$, $0.009[\dot4\dot3]$, $0.020[\dot4\dot4]$, $0.071[\dot4\dot7]$, $0.020[\dot5\dot1]$, $0.010[\dot5\dot2]$, $0.034[\dot5\dot3]$, $0.032[\dot5\dot5]$, $0.010[\dot5\dot6]$, $0.030[\dot5\dot7]$, $0.022[\dot5\dot8]$, $0.234[\dot6\dot1]$, $0.100[\dot6\dot2]$, $0.002[\dot6\dot3]$, $0.011[\dot6\dot4]$, $0.783[\dot6\dot5]$, $0.017[\dot6\dot6]$, $0.069[\dot6\dot7]$, $0.068[\dot7\dot1]$, $0.058[\dot7\dot2]$, $0.022[\dot7\dot4]$, $0.077[\dot7\dot5]$, $0.037[\dot7\dot6]$, $0.148[\dot8\dot1]$, $0.001[\dot8\dot3]$, $0.806[\dot8\dot4]$, $0.010[\dot8\dot5]$, $0.113[\dot8\dot7]$; $\Delta=0.003190$}\vskip1ex}\end{minipage}\\
\hline
\end{longtable}
\smallbreak
\begin{longtable}{|l|}\hline
$R=2.591\quad \Lambda=-85.757\quad {\rm SUSY}=(0,0)\quad {\rm dim}(H)=1$\\\hline
\hbox to3.5em{$\Psi_{128}$} \begin{minipage}{24em}{\tiny\vskip1ex{$0.001[14]$, $0.232[18]$, $0.145[22]$, $0.007[23]$, $0.083[26]$, $0.005[27]$, $0.001[33]$, $0.008[42]$, $0.036[46]$, $0.029[48]$, $0.002[51]$, $0.001[53]$, $0.029[61]$, $0.102[62]$, $0.001[67]$, $0.009[68]$, $0.232[71]$, $0.001[74]$, $0.002[81]$, $0.044[82]$, $0.084[83]$, $0.928[\dot1\dot8]$, $0.083[\dot2\dot1]$, $0.007[\dot2\dot8]$, $0.036[\dot4\dot1]$, $0.029[\dot4\dot3]$, $0.013[\dot4\dot7]$, $0.003[\dot5\dot6]$, $0.009[\dot6\dot3]$, $0.029[\dot6\dot6]$, $0.928[\dot7\dot1]$, $0.002[\dot8\dot6]$, $0.018[\dot8\dot7]$, $0.084[\dot8\dot8]$; $\Delta=0.002968$}\vskip1ex}\end{minipage}\\\hline
\hbox to3.5em{$\mathcal{M}_s/g^2$} \begin{minipage}{24em}{\tiny\vskip1ex{$-110.537{(\times 2)}$, $ -85.154$, $ -81.972{(\times 2)}$, $ -81.125{(\times 2)}$, $ -81.012{(\times 2)}$, $ -80.548$, $ -79.019{(\times 2)}$, $ -78.380{(\times 2)}$, $ -76.828{(\times 2)}$, $ -72.632$, $ -72.563$, $ -71.140$, $ -71.093$, $ -71.092$, $ -69.317{(\times 2)}$, $ -67.728{(\times 2)}$, $ -66.981{(\times 2)}$, $ -66.871$, $ -66.515{(\times 2)}$, $ -56.396{(\times 2)}$, $ -54.408{(\times 2)}$, $ -51.967{(\times 2)}$, $ -51.263{(\times 2)}$, $ -42.318{(\times 2)}$, $ -36.981{(\times 2)}$, $ -35.868$, $ -34.710{(\times 2)}$, $ -34.591{(\times 2)}$, $ -30.358$, $ -25.177$, $ -24.350{(\times 2)}$, $ -23.754$, $ -16.549{(\times 2)}$, $   0.000{(\times 55)}$, $   7.125{(\times 2)}$, $  19.115$, $  27.604$, $ 128.434$, $ 144.288{(\times 2)}$, $ 166.525{(\times 2)}$, $ 197.294$, $ 219.723$, $ 219.728$, $ 254.717$, $ 258.804{(\times 2)}$, $ 340.023$, $ 340.024$, $ 354.055$, $ 400.532$}\vskip1ex}\end{minipage}\\
\hline
\end{longtable}
\smallbreak
\begin{longtable}{|l|}\hline
$R=2.648\quad \Lambda=-93.255\quad {\rm SUSY}=(1,1)\quad {\rm dim}(H)=2$\\\hline
\hbox to3.5em{$\Psi_{128}$} \begin{minipage}{24em}{\tiny\vskip1ex{$0.875[51]$, $0.116[67]$, $0.875[76]$, $0.965[84]$, $0.875[\dot2\dot3]$, $0.965[\dot3\dot2]$, $0.205[\dot5\dot4]$, $0.116[\dot8\dot1]$; $\Delta=0.001631$}\vskip1ex}\end{minipage}\\\hline
\hbox to3.5em{$\mathcal{M}_s/g^2$} \begin{minipage}{24em}{\tiny\vskip1ex{$ -93.255{(\times 3)}$, $ -91.255{(\times 8)}$, $ -85.255{(\times 5)}$, $ -81.598{(\times 8)}$, $ -81.172{(\times 2)}$, $ -70.608{(\times 2)}$, $ -69.264{(\times 2)}$, $ -57.436{(\times 4)}$, $ -55.052{(\times 2)}$, $ -54.284{(\times 8)}$, $ -46.627{(\times 6)}$, $ -35.818{(\times 4)}$, $   0.000{(\times 55)}$, $  12.927{(\times 2)}$, $  44.255{(\times 2)}$, $  93.255$, $ 133.881{(\times 4)}$, $ 133.883{(\times 4)}$, $ 178.509$, $ 305.198{(\times 2)}$, $ 363.990{(\times 2)}$, $ 450.273$}\vskip1ex}\end{minipage}\\\hline
\hbox to3.5em{$\mathcal{M}_v/g$} \begin{minipage}{24em}{\tiny\vskip1ex{$ -16.485{(\times 3)}$, $ -15.071{(\times 4)}$, $ -13.071{(\times 4)}$, $  -7.579{(\times 2)}$, $  -6.828{(\times 3)}$, $  -5.985{(\times 2)}$, $  -3.414{(\times 4)}$, $  -1.414{(\times 4)}$, $   0.000{(\times 76)}$, $   1.414{(\times 4)}$, $   3.414{(\times 4)}$, $   5.985{(\times 2)}$, $   6.828{(\times 3)}$, $   7.579{(\times 2)}$, $  13.071{(\times 4)}$, $  15.071{(\times 4)}$, $  16.485{(\times 3)}$}\vskip1ex}\end{minipage}\\\hline
\hbox to3.5em{$A_1$} \begin{minipage}{24em}{\tiny\vskip1ex{$ -11.657{(\times 3)}$, $  -8.243{(\times 4)}$, $  -4.828$, $   4.828$, $   8.243{(\times 4)}$, $  11.657{(\times 3)}$}\vskip1ex}\end{minipage}\\
\hline
\end{longtable}
\smallbreak
\begin{longtable}{|l|}\hline
$R=2.649\quad \Lambda=-90.252\quad {\rm SUSY}=(0,0)\quad {\rm dim}(H)=0$\\\hline
\hbox to3.5em{$\Psi_{128}$} \begin{minipage}{24em}{\tiny\vskip1ex{$0.004[12]$, $0.006[13]$, $0.102[14]$, $0.022[15]$, $0.001[16]$, $0.002[22]$, $0.131[25]$, $0.003[26]$, $0.009[27]$, $0.122[28]$, $0.004[31]$, $0.019[32]$, $0.004[34]$, $0.008[35]$, $0.050[36]$, $0.129[37]$, $0.013[42]$, $0.004[44]$, $0.015[45]$, $0.247[46]$, $0.012[48]$, $0.005[51]$, $0.119[52]$, $0.004[54]$, $0.740[56]$, $0.058[57]$, $0.039[62]$, $0.006[64]$, $0.003[65]$, $0.004[71]$, $0.001[73]$, $0.174[74]$, $0.723[75]$, $0.001[76]$, $0.027[82]$, $0.009[83]$, $0.023[85]$, $0.004[87]$, $0.159[88]$, $0.005[\dot1\dot2]$, $0.002[\dot1\dot3]$, $0.002[\dot1\dot6]$, $0.003[\dot1\dot7]$, $0.013[\dot2\dot1]$, $0.001[\dot2\dot2]$, $1.201[\dot2\dot4]$, $0.002[\dot3\dot4]$, $0.029[\dot3\dot6]$, $1.138[\dot3\dot7]$, $0.001[\dot3\dot8]$, $0.001[\dot4\dot1]$, $0.024[\dot4\dot2]$, $0.010[\dot4\dot3]$, $0.002[\dot4\dot4]$, $0.003[\dot4\dot5]$, $0.003[\dot4\dot6]$, $0.012[\dot4\dot7]$, $0.002[\dot4\dot8]$, $0.005[\dot5\dot4]$, $0.005[\dot5\dot5]$, $1.191[\dot5\dot6]$, $0.026[\dot5\dot7]$, $0.001[\dot5\dot8]$, $0.028[\dot6\dot7]$, $0.080[\dot7\dot4]$, $0.052[\dot7\dot5]$, $0.002[\dot7\dot7]$, $0.015[\dot7\dot8]$, $0.004[\dot8\dot2]$, $0.002[\dot8\dot3]$, $0.027[\dot8\dot4]$, $0.019[\dot8\dot5]$, $0.022[\dot8\dot8]$; $\Delta=0.003074$}\vskip1ex}\end{minipage}\\
\hline
\end{longtable}
\smallbreak
\begin{longtable}{|l|}\hline
$R=2.664\quad \Lambda=-94.136\quad {\rm SUSY}=(0,0)\quad {\rm dim}(H)=0$\\
\hline
\end{longtable}
\smallbreak
\begin{longtable}{|l|}\hline
$R=2.686\quad \Lambda=-95.538\quad {\rm SUSY}=(0,0)\quad {\rm dim}(H)=0$\\\hline
\hbox to3.5em{$\Psi_{128}$} \begin{minipage}{24em}{\tiny\vskip1ex{$0.044[11]$, $0.033[15]$, $0.018[16]$, $0.998[21]$, $0.012[24]$, $0.005[26]$, $0.023[27]$, $0.007[31]$, $0.197[35]$, $0.185[36]$, $0.005[37]$, $0.048[38]$, $0.001[41]$, $0.012[42]$, $0.154[44]$, $0.074[46]$, $0.004[51]$, $0.168[53]$, $0.088[54]$, $0.003[57]$, $0.009[62]$, $0.999[63]$, $0.054[64]$, $0.045[65]$, $0.001[67]$, $0.007[68]$, $0.046[72]$, $0.005[73]$, $0.193[77]$, $0.060[78]$, $0.074[81]$, $0.002[83]$, $0.011[84]$, $0.001[85]$, $0.004[86]$, $1.008[87]$, $0.047[\dot1\dot2]$, $0.006[\dot1\dot3]$, $0.024[\dot1\dot7]$, $0.998[\dot1\dot8]$, $0.065[\dot2\dot2]$, $0.016[\dot2\dot3]$, $0.053[\dot2\dot4]$, $0.032[\dot2\dot5]$, $0.044[\dot2\dot8]$, $0.018[\dot3\dot1]$, $0.012[\dot3\dot2]$, $0.154[\dot3\dot4]$, $0.086[\dot3\dot5]$, $0.060[\dot3\dot6]$, $0.006[\dot3\dot7]$, $0.001[\dot4\dot2]$, $0.057[\dot4\dot6]$, $0.005[\dot5\dot1]$, $0.379[\dot5\dot3]$, $0.088[\dot5\dot4]$, $0.045[\dot5\dot5]$, $0.006[\dot6\dot1]$, $0.045[\dot6\dot5]$, $0.999[\dot6\dot6]$, $0.060[\dot7\dot1]$, $0.009[\dot7\dot4]$, $0.003[\dot7\dot6]$, $0.193[\dot7\dot7]$, $0.005[\dot8\dot1]$, $0.011[\dot8\dot4]$; $\Delta=0.003249$}\vskip1ex}\end{minipage}\\
\hline
\end{longtable}
\smallbreak
\begin{longtable}{|l|}\hline
$R=2.691\quad \Lambda=-94.399\quad {\rm SUSY}=(0,0)\quad {\rm dim}(H)=0$\\\hline
\hbox to3.5em{$\Psi_{128}$} \begin{minipage}{24em}{\tiny\vskip1ex{$0.011[11]$, $0.187[14]$, $0.055[36]$, $0.001[51]$, $0.034[55]$, $0.064[58]$, $0.034[68]$, $0.014[72]$, $0.090[73]$, $1.099[83]$, $0.207[\dot1\dot6]$, $0.260[\dot3\dot1]$, $0.047[\dot3\dot4]$, $0.047[\dot4\dot1]$, $0.509[\dot4\dot4]$, $0.030[\dot6\dot2]$, $0.001[\dot7\dot4]$, $0.279[\dot7\dot5]$, $0.533[\dot8\dot5]$; $\Delta=0.002083$}\vskip1ex}\end{minipage}\\\hline
\hbox to3.5em{$\mathcal{M}_s/g^2$} \begin{minipage}{24em}{\tiny\vskip1ex{$-121.785$, $ -98.755{(\times 2)}$, $ -97.579$, $ -93.792$, $ -91.882{(\times 2)}$, $ -89.540$, $ -88.136$, $ -83.853{(\times 2)}$, $ -83.774$, $ -82.712$, $ -80.961$, $ -79.804{(\times 2)}$, $ -77.442$, $ -75.492{(\times 2)}$, $ -70.424$, $ -64.680{(\times 2)}$, $ -64.035$, $ -62.120$, $ -61.485$, $ -61.101{(\times 2)}$, $ -59.834{(\times 2)}$, $ -59.595$, $ -58.384$, $ -55.015$, $ -54.961$, $ -52.921$, $ -48.437{(\times 2)}$, $ -44.195$, $ -41.897$, $ -36.390{(\times 2)}$, $ -31.399$, $ -28.347{(\times 2)}$, $ -27.610$, $ -27.567$, $ -25.228$, $ -24.399$, $ -23.262{(\times 2)}$, $ -19.335$, $ -13.100$, $   0.000{(\times 56)}$, $  30.578$, $  40.534$, $  87.354$, $  96.692$, $ 111.696$, $ 131.478{(\times 2)}$, $ 165.365{(\times 2)}$, $ 217.116$, $ 252.738$, $ 259.934$, $ 280.677$, $ 301.488{(\times 2)}$, $ 356.755{(\times 2)}$, $ 364.395$, $ 414.906$, $ 415.620$, $ 434.635$}\vskip1ex}\end{minipage}\\
\hline
\end{longtable}
\smallbreak
\begin{longtable}{|l|}\hline
$R=2.694\quad \Lambda=-94.909\quad {\rm SUSY}=(0,0)\quad {\rm dim}(H)=0$\\\hline
\hbox to3.5em{$\Psi_{128}$} \begin{minipage}{24em}{\tiny\vskip1ex{$-0.183[15]$, $0.030[17]$, $0.914[26]$, $-0.183[28]$, $0.000[32]$, $-0.183[36]$, $-0.030[38]$, $-0.914[45]$, $-0.183[47]$, $-0.274[52]$, $-0.210[54]$, $0.000[58]$, $0.717[61]$, $0.232[63]$, $0.232[71]$, $0.717[73]$, $-0.210[82]$, $-0.274[84]$, $0.000[88]$, $0.089[\dot1\dot5]$, $0.007[\dot1\dot7]$, $-0.007[\dot2\dot6]$, $-0.089[\dot2\dot8]$, $-0.089[\dot3\dot6]$, $1.148[\dot3\dot8]$, $-1.148[\dot4\dot5]$, $0.089[\dot4\dot7]$, $-0.189[\dot5\dot2]$, $-0.189[\dot6\dot3]$, $1.155[\dot7\dot1]$, $-0.014[\dot8\dot4]$; $\Delta=0.001376$}\vskip1ex}\end{minipage}\\\hline
\hbox to3.5em{$\mathcal{M}_s/g^2$} \begin{minipage}{24em}{\tiny\vskip1ex{$-122.159$, $-101.902{(\times 3)}$, $ -99.564{(\times 3)}$, $ -92.320$, $ -91.114{(\times 3)}$, $ -89.594$, $ -84.679{(\times 3)}$, $ -81.085{(\times 3)}$, $ -67.603{(\times 2)}$, $ -65.087{(\times 3)}$, $ -60.114{(\times 2)}$, $ -57.070{(\times 3)}$, $ -56.991$, $ -56.302{(\times 3)}$, $ -51.834$, $ -51.633{(\times 2)}$, $ -48.853{(\times 3)}$, $ -34.254{(\times 3)}$, $ -29.201$, $ -29.200{(\times 2)}$, $ -24.829{(\times 2)}$, $ -24.828$, $ -22.066{(\times 2)}$, $ -15.366$, $   0.000{(\times 56)}$, $  26.021$, $  44.325{(\times 2)}$, $  75.286{(\times 2)}$, $ 118.421$, $ 118.422{(\times 2)}$, $ 188.957{(\times 3)}$, $ 219.164{(\times 2)}$, $ 272.548{(\times 2)}$, $ 272.549$, $ 390.285{(\times 2)}$, $ 390.286$, $ 399.404{(\times 2)}$, $ 419.084$}\vskip1ex}\end{minipage}\\
\hline
\end{longtable}
\smallbreak
\begin{longtable}{|l|}\hline
$R=2.702\quad \Lambda=-98.908\quad {\rm SUSY}=(0,0)\quad {\rm dim}(H)=2$\\\hline
\hbox to3.5em{$\Psi_{128}$} \begin{minipage}{24em}{\tiny\vskip1ex{$-0.000[11]$, $1.112[32]$, $-0.187[41]$, $-0.894[54]$, $-0.000[57]$, $0.587[67]$, $0.000[73]$, $-0.894[78]$, $0.587[83]$, $0.006[\dot1\dot3]$, $-0.104[\dot1\dot4]$, $0.104[\dot1\dot7]$, $0.006[\dot1\dot8]$, $0.104[\dot2\dot3]$, $-0.006[\dot2\dot4]$, $0.006[\dot2\dot7]$, $0.104[\dot2\dot8]$, $0.104[\dot3\dot3]$, $-0.006[\dot3\dot4]$, $0.006[\dot3\dot7]$, $0.104[\dot3\dot8]$, $-0.006[\dot4\dot3]$, $0.104[\dot4\dot4]$, $-0.104[\dot4\dot7]$, $-0.006[\dot4\dot8]$, $0.954[\dot5\dot2]$, $-0.055[\dot5\dot5]$, $0.055[\dot6\dot1]$, $-0.954[\dot6\dot6]$, $0.954[\dot7\dot1]$, $-0.055[\dot7\dot6]$, $0.055[\dot8\dot2]$, $-0.954[\dot8\dot5]$; $\Delta=0.001723$}\vskip1ex}\end{minipage}\\\hline
\hbox to3.5em{$\mathcal{M}_s/g^2$} \begin{minipage}{24em}{\tiny\vskip1ex{$-114.284{(\times 2)}$, $-108.877$, $-103.444{(\times 4)}$, $ -94.504{(\times 2)}$, $ -91.753$, $ -86.033{(\times 2)}$, $ -85.010{(\times 4)}$, $ -82.711$, $ -82.710$, $ -79.578{(\times 2)}$, $ -79.126{(\times 4)}$, $ -75.048{(\times 2)}$, $ -72.675{(\times 2)}$, $ -72.247$, $ -64.016$, $ -60.598{(\times 2)}$, $ -58.121$, $ -58.119{(\times 2)}$, $ -56.128{(\times 4)}$, $ -56.109{(\times 2)}$, $ -43.410{(\times 2)}$, $ -43.027{(\times 2)}$, $ -39.641{(\times 4)}$, $ -35.550$, $ -28.480{(\times 2)}$, $ -16.709{(\times 2)}$, $   0.000{(\times 54)}$, $   3.514$, $  10.770{(\times 2)}$, $  26.349{(\times 2)}$, $ 107.681{(\times 2)}$, $ 146.605$, $ 187.706$, $ 187.707{(\times 3)}$, $ 239.827{(\times 2)}$, $ 270.100{(\times 2)}$, $ 287.283{(\times 2)}$, $ 363.550$, $ 480.910$}\vskip1ex}\end{minipage}\\
\hline
\end{longtable}
\smallbreak
\begin{longtable}{|l|}\hline
$R=2.714\quad \Lambda=-97.155\quad {\rm SUSY}=(0,0)\quad {\rm dim}(H)=0 [?]$\\\hline
\hbox to3.5em{$\Psi_{128}$} \begin{minipage}{24em}{\tiny\vskip1ex{$0.028[13]$, $0.025[14]$, $0.001[15]$, $0.002[16]$, $1.171[17]$, $0.001[21]$, $0.032[23]$, $0.003[26]$, $0.044[27]$, $0.004[31]$, $0.002[32]$, $0.032[34]$, $0.062[37]$, $0.019[38]$, $0.005[41]$, $0.041[47]$, $0.025[52]$, $0.002[53]$, $0.004[54]$, $0.001[57]$, $1.188[62]$, $0.002[64]$, $0.015[65]$, $0.070[66]$, $0.001[67]$, $0.036[75]$, $0.059[76]$, $0.002[77]$, $0.004[78]$, $0.004[84]$, $0.035[86]$, $0.003[87]$, $0.336[\dot1\dot4]$, $0.015[\dot1\dot5]$, $0.009[\dot1\dot7]$, $0.005[\dot2\dot1]$, $0.007[\dot2\dot2]$, $0.312[\dot2\dot4]$, $0.027[\dot2\dot5]$, $0.025[\dot2\dot7]$, $0.130[\dot3\dot3]$, $0.672[\dot3\dot4]$, $0.021[\dot3\dot5]$, $0.133[\dot3\dot7]$, $0.246[\dot5\dot5]$, $0.026[\dot5\dot6]$, $0.005[\dot5\dot7]$, $0.011[\dot5\dot8]$, $0.050[\dot6\dot2]$, $0.015[\dot6\dot3]$, $0.294[\dot6\dot5]$, $0.005[\dot6\dot8]$, $0.036[\dot7\dot1]$, $0.177[\dot7\dot2]$, $0.008[\dot7\dot3]$, $0.005[\dot7\dot4]$, $0.626[\dot7\dot5]$, $0.134[\dot7\dot6]$, $0.004[\dot7\dot7]$, $0.006[\dot7\dot8]$, $0.035[\dot8\dot1]$, $0.736[\dot8\dot2]$, $0.027[\dot8\dot3]$, $0.036[\dot8\dot5]$; $\Delta=0.003327$}\vskip1ex}\end{minipage}\\
\hline
\end{longtable}
\smallbreak
\begin{longtable}{|l|}\hline
$R=2.765\quad \Lambda=-105.528\quad {\rm SUSY}=(1,1)\quad {\rm dim}(H)=8$\\\hline
\hbox to3.5em{$\Psi_{128}$} \begin{minipage}{24em}{\tiny\vskip1ex{$0.430[\dot1\dot1]$, $0.430[\dot2\dot2]$, $0.037[\dot3\dot3]$, $1.164[\dot3\dot5]$, $0.430[\dot4\dot4]$, $1.164[\dot5\dot3]$, $0.037[\dot5\dot5]$, $0.430[\dot6\dot6]$, $0.430[\dot7\dot7]$, $0.430[\dot8\dot8]$; $\Delta=0.001601$}\vskip1ex}\end{minipage}\\\hline
\hbox to3.5em{$\mathcal{M}_s/g^2$} \begin{minipage}{24em}{\tiny\vskip1ex{$-105.528{(\times 8)}$, $-104.836{(\times 12)}$, $ -99.646$, $ -84.488{(\times 6)}$, $ -81.968{(\times 8)}$, $ -73.200{(\times 6)}$, $ -43.945$, $ -16.399{(\times 12)}$, $   0.000{(\times 48)}$, $  93.725$, $ 123.284{(\times 8)}$, $ 168.195{(\times 12)}$, $ 220.267$, $ 235.669$, $ 489.263$, $ 696.633$, $ 720.700$}\vskip1ex}\end{minipage}\\\hline
\hbox to3.5em{$\mathcal{M}_v/g$} \begin{minipage}{24em}{\tiny\vskip1ex{$ -18.050$, $ -16.545{(\times 6)}$, $ -15.958{(\times 6)}$, $  -7.777$, $  -5.686{(\times 6)}$, $   0.000{(\times 88)}$, $   5.686{(\times 6)}$, $   7.777$, $  15.958{(\times 6)}$, $  16.545{(\times 6)}$, $  18.050$}\vskip1ex}\end{minipage}\\\hline
\hbox to3.5em{$A_1$} \begin{minipage}{24em}{\tiny\vskip1ex{$ -12.913$, $ -10.822{(\times 6)}$, $  -5.136$, $   5.136$, $  10.822{(\times 6)}$, $  12.913$}\vskip1ex}\end{minipage}\\\hline
\hbox to3.5em{$A_3$} \begin{minipage}{24em}{\tiny\vskip1ex{$ -38.740$, $ -32.466{(\times 6)}$, $ -23.608$, $ -23.186$, $ -21.681{(\times 6)}$, $ -19.252$, $ -12.913$, $ -11.408{(\times 6)}$, $ -10.822{(\times 6)}$, $  -9.990{(\times 8)}$, $  -5.136{(\times 8)}$, $  -4.304{(\times 12)}$, $  -2.711$, $  -0.549{(\times 6)}$, $   0.549{(\times 6)}$, $   2.711$, $   4.304{(\times 12)}$, $   5.136{(\times 8)}$, $   9.990{(\times 8)}$, $  10.822{(\times 6)}$, $  11.408{(\times 6)}$, $  12.913$, $  19.252$, $  21.681{(\times 6)}$, $  23.186$, $  23.608$, $  32.466{(\times 6)}$, $  38.740$}\vskip1ex}\end{minipage}\\\hline
\end{longtable}
\smallbreak
\begin{longtable}{|l|}\hline
$R=2.794\quad \Lambda=-110.183\quad {\rm SUSY}=(0,0)\quad {\rm dim}(H)=0$\\\hline
\hbox to3.5em{$\Psi_{128}$} \begin{minipage}{24em}{\tiny\vskip1ex{$0.001[15]$, $0.007[16]$, $0.382[17]$, $0.012[22]$, $1.103[23]$, $0.001[26]$, $0.940[31]$, $0.050[41]$, $0.001[44]$, $0.001[47]$, $0.382[54]$, $0.009[56]$, $0.250[58]$, $0.001[61]$, $0.250[64]$, $0.001[66]$, $0.123[68]$, $0.187[71]$, $0.153[75]$, $0.001[78]$, $0.153[81]$, $0.001[85]$, $0.153[86]$, $0.001[87]$, $0.038[\dot1\dot4]$, $0.003[\dot1\dot6]$, $0.038[\dot1\dot7]$, $0.002[\dot2\dot2]$, $0.054[\dot2\dot4]$, $0.001[\dot2\dot5]$, $0.009[\dot2\dot6]$, $0.953[\dot2\dot7]$, $0.138[\dot3\dot1]$, $0.006[\dot4\dot1]$, $0.138[\dot4\dot2]$, $0.006[\dot4\dot4]$, $0.013[\dot4\dot5]$, $0.052[\dot4\dot6]$, $0.009[\dot4\dot7]$, $0.072[\dot5\dot7]$, $0.898[\dot5\dot8]$, $0.002[\dot6\dot2]$, $0.953[\dot6\dot4]$, $0.006[\dot6\dot6]$, $0.054[\dot6\dot7]$, $0.072[\dot6\dot8]$, $0.013[\dot7\dot1]$, $0.013[\dot7\dot6]$, $0.001[\dot7\dot7]$, $0.001[\dot7\dot8]$, $0.138[\dot8\dot2]$, $0.009[\dot8\dot4]$, $0.006[\dot8\dot7]$; $\Delta=0.002903$}\vskip1ex}\end{minipage}\\
\hline
\end{longtable}
\smallbreak
\begin{longtable}{|l|}\hline
$R=2.930\quad \Lambda=-129.687\quad {\rm SUSY}=(0,0)\quad {\rm dim}(H)=0$\\\hline
\hbox to3.5em{$\Psi_{128}$} \begin{minipage}{24em}{\tiny\vskip1ex{$0.101[13]$, $0.009[14]$, $0.003[21]$, $0.149[22]$, $0.033[23]$, $0.190[24]$, $0.276[27]$, $0.057[28]$, $0.007[33]$, $0.091[36]$, $0.003[43]$, $0.058[44]$, $0.009[45]$, $0.230[46]$, $0.036[52]$, $0.186[57]$, $0.226[62]$, $0.107[64]$, $0.142[67]$, $0.093[68]$, $0.023[72]$, $0.008[73]$, $0.471[74]$, $0.413[75]$, $0.060[78]$, $0.096[81]$, $0.064[84]$, $0.008[85]$, $0.001[88]$, $0.009[\dot1\dot1]$, $0.011[\dot1\dot3]$, $0.005[\dot1\dot4]$, $0.049[\dot1\dot6]$, $0.017[\dot1\dot7]$, $0.030[\dot1\dot8]$, $0.031[\dot2\dot2]$, $0.068[\dot2\dot4]$, $0.041[\dot2\dot5]$, $0.042[\dot2\dot7]$, $0.020[\dot3\dot3]$, $0.060[\dot3\dot4]$, $0.023[\dot3\dot5]$, $0.066[\dot3\dot8]$, $0.049[\dot4\dot2]$, $0.065[\dot4\dot3]$, $0.029[\dot4\dot4]$, $1.079[\dot4\dot6]$, $0.013[\dot4\dot7]$, $0.075[\dot5\dot1]$, $0.118[\dot5\dot4]$, $0.047[\dot5\dot5]$, $0.004[\dot5\dot6]$, $0.111[\dot5\dot7]$, $0.012[\dot6\dot3]$, $0.011[\dot6\dot7]$, $1.133[\dot7\dot1]$, $0.032[\dot7\dot2]$, $0.010[\dot7\dot3]$, $0.017[\dot7\dot4]$, $0.018[\dot7\dot7]$, $0.031[\dot8\dot2]$, $0.079[\dot8\dot3]$, $0.048[\dot8\dot7]$, $0.160[\dot8\dot8]$; $\Delta=0.003224$}\vskip1ex}\end{minipage}\\
\hline
\end{longtable}
\smallbreak
\begin{longtable}{|l|}\hline
$R=2.940\quad \Lambda=-131.996\quad {\rm SUSY}=(0,0)\quad {\rm dim}(H)=0$\\\hline
\hbox to3.5em{$\Psi_{128}$} \begin{minipage}{24em}{\tiny\vskip1ex{$0.009[15]$, $0.058[16]$, $0.468[17]$, $0.057[21]$, $0.449[23]$, $0.078[27]$, $0.301[33]$, $0.001[35]$, $0.104[36]$, $0.065[44]$, $1.023[46]$, $0.009[48]$, $0.415[51]$, $0.301[54]$, $0.104[55]$, $0.001[56]$, $0.017[64]$, $0.058[68]$, $0.134[71]$, $0.025[72]$, $0.058[75]$, $0.009[76]$, $0.468[78]$, $0.301[81]$, $0.449[84]$, $0.078[88]$, $0.079[\dot1\dot4]$, $0.208[\dot1\dot8]$, $0.020[\dot2\dot5]$, $0.044[\dot3\dot1]$, $0.001[\dot3\dot6]$, $0.034[\dot3\dot7]$, $0.020[\dot4\dot4]$, $0.922[\dot4\dot5]$, $0.079[\dot4\dot8]$, $1.030[\dot5\dot2]$, $0.069[\dot5\dot8]$, $0.001[\dot6\dot1]$, $0.054[\dot6\dot3]$, $0.132[\dot6\dot6]$, $0.033[\dot6\dot7]$, $0.034[\dot7\dot1]$, $0.033[\dot7\dot6]$, $0.026[\dot8\dot1]$, $0.033[\dot8\dot3]$, $0.053[\dot8\dot7]$; $\Delta=0.002539$}\vskip1ex}\end{minipage}\\\hline
\hbox to3.5em{$\mathcal{M}_s/g^2$} \begin{minipage}{24em}{\tiny\vskip1ex{$-167.789$, $-160.012$, $-152.037$, $-151.218$, $-146.804$, $-142.342$, $-141.392$, $-129.648$, $-127.458$, $-124.191$, $-121.761$, $-120.720$, $-113.697$, $-111.125$, $-108.256$, $-107.565$, $-107.203$, $-105.111$, $ -96.798$, $ -96.515$, $ -95.424$, $ -94.499$, $ -90.414$, $ -86.739$, $ -84.720$, $ -74.863$, $ -73.257$, $ -72.680$, $ -69.820$, $ -68.793$, $ -64.039$, $ -63.079$, $ -61.121$, $ -60.869$, $ -57.972$, $ -57.434$, $ -53.590$, $ -52.779$, $ -52.605$, $ -40.282$, $ -39.977$, $ -30.601$, $ -18.838$, $ -10.241$, $  -2.589$, $   0.000{(\times 56)}$, $  30.876$, $  32.011$, $  35.794$, $  47.087$, $  81.767$, $ 102.099$, $ 105.940$, $ 152.854$, $ 156.832$, $ 191.620$, $ 217.279$, $ 270.901$, $ 289.097$, $ 294.233$, $ 303.499$, $ 310.195$, $ 311.218$, $ 318.183$, $ 360.771$, $ 372.630$, $ 400.411$, $ 487.363$, $ 533.051$, $ 561.596$, $ 576.078$, $ 617.812$, $ 630.527$}\vskip1ex}\end{minipage}\\
\hline
\end{longtable}
\smallbreak
\begin{longtable}{|l|}\hline
$R=2.994\quad \Lambda=-139.990\quad {\rm SUSY}=(0,0)\quad {\rm dim}(H)=0$\\
\hline
\end{longtable}
\smallbreak
\input{vacua/R3073.vac.tex}
\smallbreak
\begin{longtable}{|l|}\hline
$R=3.195\quad \Lambda=-186.317\quad {\rm SUSY}=(0,0)\quad {\rm dim}(H)=0$\\\hline
\hbox to3.5em{$\Psi_{128}$} \begin{minipage}{24em}{\tiny\vskip1ex{$0.073[13]$, $0.041[18]$, $0.027[23]$, $0.046[24]$, $0.007[27]$, $0.076[32]$, $0.904[34]$, $0.026[35]$, $0.065[36]$, $0.060[37]$, $0.011[38]$, $0.002[42]$, $0.019[45]$, $0.090[51]$, $0.151[52]$, $0.011[53]$, $0.078[54]$, $0.055[56]$, $0.054[61]$, $0.152[62]$, $0.034[64]$, $0.039[67]$, $0.039[68]$, $0.004[72]$, $0.018[74]$, $0.145[75]$, $0.186[76]$, $0.046[84]$, $0.085[85]$, $0.128[87]$, $0.014[88]$, $0.023[\dot1\dot1]$, $0.047[\dot1\dot2]$, $0.067[\dot1\dot3]$, $0.188[\dot1\dot6]$, $0.069[\dot2\dot1]$, $0.170[\dot2\dot6]$, $0.084[\dot2\dot7]$, $0.003[\dot3\dot1]$, $0.019[\dot3\dot5]$, $0.006[\dot3\dot6]$, $0.914[\dot3\dot7]$, $0.029[\dot3\dot8]$, $0.046[\dot4\dot3]$, $0.036[\dot4\dot4]$, $0.030[\dot4\dot6]$, $0.044[\dot4\dot7]$, $0.187[\dot5\dot2]$, $0.285[\dot5\dot3]$, $0.064[\dot5\dot4]$, $0.050[\dot6\dot1]$, $1.035[\dot6\dot5]$, $0.018[\dot6\dot7]$, $0.041[\dot6\dot8]$, $1.129[\dot7\dot1]$, $0.018[\dot7\dot8]$, $0.194[\dot8\dot6]$, $0.147[\dot8\dot8]$; $\Delta=0.003310$}\vskip1ex}\end{minipage}\\
\hline
\end{longtable}
\smallbreak
\input{vacua/R3212.vac.tex}
\smallbreak
\input{vacua/R3323.vac.tex}
\smallbreak
\input{vacua/R3408.vac.tex}
\smallbreak
\input{vacua/R3419.vac.tex}
\smallbreak
\input{vacua/R3452.vac.tex}
\smallbreak
\input{vacua/R3463.vac.tex}
\smallbreak
\input{vacua/R3508.vac.tex}
\smallbreak
\input{vacua/R3518.vac.tex}
\smallbreak
\input{vacua/R3525.vac.tex}
\smallbreak
\input{vacua/R3541.vac.tex}
\smallbreak
\input{vacua/R3542.vac.tex}
\smallbreak
\input{vacua/R3545.vac.tex}
\smallbreak
\input{vacua/R3554.vac.tex}
\smallbreak
\input{vacua/R3555.vac.tex}
\smallbreak
\input{vacua/R3566.vac.tex}
\smallbreak
\input{vacua/R3575.vac.tex}
\smallbreak
\input{vacua/R3593.vac.tex}
\smallbreak
\input{vacua/R3598.vac.tex}
\smallbreak
\input{vacua/R3610.vac.tex}
\smallbreak
\input{vacua/R3653.vac.tex}
\smallbreak
\input{vacua/R3665.vac.tex}
\smallbreak
\input{vacua/R3666.vac.tex}
\smallbreak
\input{vacua/R3703.vac.tex}
\smallbreak
\input{vacua/R3770.vac.tex}
\smallbreak
\input{vacua/R3785.vac.tex}
\smallbreak
\input{vacua/R3790.vac.tex}
\smallbreak
\input{vacua/R3794.vac.tex}
\smallbreak
\input{vacua/R3839.vac.tex}
\smallbreak
\input{vacua/R3869.vac.tex}
\smallbreak
\input{vacua/R3874.vac.tex}
\smallbreak
\input{vacua/R3928.vac.tex}
\smallbreak
\input{vacua/R4002.vac.tex}
\smallbreak
\input{vacua/R4032.vac.tex}
\smallbreak
\input{vacua/R4050.vac.tex}
\smallbreak
\input{vacua/R4075.vac.tex}
\smallbreak
\input{vacua/R4094.vac.tex}
\smallbreak
\input{vacua/R4190.vac.tex}
\smallbreak
\input{vacua/R4309.vac.tex}
\smallbreak
\input{vacua/R4357.vac.tex}
\smallbreak
\input{vacua/R4377.vac.tex}
\smallbreak
\input{vacua/R4501.vac.tex}
\smallbreak
\input{vacua/R4509.vac.tex}
\smallbreak
\input{vacua/R4521.vac.tex}
\smallbreak
\input{vacua/R4528.vac.tex}
\smallbreak
\input{vacua/R4541.vac.tex}
\smallbreak
\input{vacua/R4627.vac.tex}
\smallbreak
\input{vacua/R4731.vac.tex}
\smallbreak
\input{vacua/R4742.vac.tex}
\smallbreak
\input{vacua/R4754.vac.tex}
\smallbreak
\input{vacua/R4763.vac.tex}
\smallbreak
\input{vacua/R4780.vac.tex}
\smallbreak
\input{vacua/R4807.vac.tex}
\smallbreak
\input{vacua/R4897.vac.tex}
\smallbreak
\input{vacua/R4970.vac.tex}
\smallbreak
\input{vacua/R4980.vac.tex}
\smallbreak
\input{vacua/R4990.vac.tex}
\smallbreak
\input{vacua/R4993.vac.tex}
\smallbreak
\input{vacua/R4994.vac.tex}
\smallbreak
\input{vacua/R5018.vac.tex}
\smallbreak
\input{vacua/R5020.vac.tex}
\smallbreak
\input{vacua/R5142.vac.tex}
\smallbreak
\input{vacua/R5330.vac.tex}
\smallbreak
\input{vacua/R5608.vac.tex}

\section{Conclusion and Outlook}

The utilization of sensitivity analysis backpropagation for the
numerical study of the scalar potentials of extended supergravity
models appears to be a dramatically successful technique for finding
and determining some of the properties of their vacua. A detailed
analysis of the one model with the technically most challenging scalar
potential known reveals that the number of possible vacuum solutions
is quite large, much larger than the number of solutions that have
been found hitherto. While some computationally helpful special
properties of Spin(16) $\Gamma$-matrices simplify the problem
in~$D=3$, it is expected that these techniques will turn out to be
equally powerful for the analysis of the~$D=4$ and~$D=5$ potentials
which involve the smaller exceptional groups~$E_7$ and~$E_6$. Detailed
information about the properties of novel vacua of these models should
be of considerable value in situations where the AdS$_4$/CFT$_3$ and
AdS$_5$/CFT$_4$ correspondences can be utilized to obtain new
insights. While a detailed analysis of the symmetry breaking structure
of~\emph{all} dimensionally reduced models of maximal supergravity
(including a large variety of noncompact gauge groups) may be within
technological reach now, special attention should hence be paid to
these particular cases.

As the number of stationary points of the scalar potential of
$SO(8)$-gauged $D=4, \cN=8$ supergravity is expected to be considerably
smaller than that of the maximal three-dimensional model analyzed
here, and as it is technically far easier to work with densely
occupied complex $56\times 56$ matrices than with real $248\times 248$
matrices, the most sensible strategy to approach this particular
problem seems to be to aim for very high numerical precision
(i.e. hundreds of digits) right from the beginning so that 
conjectures for analytic expressions for the locations and properties of
stationary points can be generated semi-automatically~\cite{Fischbacher_in_progress},
to serve as important guidance for stringent proofs establishing the 
existence of new solutions.

\paragraph{Acknowledgments}

It is a pleasure to thank Hermann Nicolai, Maria Alessandra Papa, and
Steffen Grunewald both for hospitality and access to supercomputing
resources at the Albert Einstein Institute where part of this work was
done. Also, I benefitted from discussions with Andreas Griewank, Shaun
Forth, and Andy Keane on automatic differentiation as well as optimization.

\end{document}